\documentclass[
 reprint,
 bibnotes,
 amsmath,
 amssymb,
 aps,
 prb,
]{revtex4-2}

\makeatletter
\AtBeginDocument{%
}
\makeatother

\newcommand{\setupSI}{
    \clearpage
 
    \setcounter{equation}{0}
    \setcounter{figure}{0}
    \setcounter{table}{0}
    \setcounter{page}{1}
    \setcounter{section}{0}

    \renewcommand{\theequation}{S\arabic{equation}}
    \renewcommand{\thefigure}{S\arabic{figure}}
    \renewcommand{\thetable}{S\arabic{table}}
    \renewcommand{\thesection}{Supplementary Note S\arabic{section}}
    
    \renewcommand{\thepage}{\arabic{page}}

    \onecolumngrid 
    \begin{center}
        \textbf{\large Supplementary Information for: }\\
        \textbf{\large Exciton-polariton condensate in the van der Waals magnet CrSBr}
    \end{center}
    \vspace{1cm}
}

\usepackage{graphicx}
\usepackage{dcolumn}
\usepackage{bm}
\usepackage{braket}

\usepackage[mathlines]{lineno}

\usepackage{xcolor}

\usepackage[T1]{fontenc}

\usepackage{hyperref}

\begin{document}

\preprint{arXiv:2501.18233v2}

\title{Exciton-polariton condensate in the van der Waals magnet CrSBr}

\author
{Bo Han,$^{1,13}$ Hangyong Shan,$^{1,13}$ Kok Wee Song,$^{2,3,13}$ Meng Sun,$^{4,13}$ Andrei Bulavin,$^{5,6}$ Ivan G. Savenko,$^{5,6,7}$  Martin Esmann,$^{1}$ Marti Struve,$^{1}$ Vita Solovyeva,$^{1}$ Lukas Lackner,$^{1}$ Falk Eilenberger,$^{8,9,10}$ Jakub Regner,$^{11}$ Zdeněk Sofer,$^{11}$ Oleksandr Kyriienko,$^{12}$ and Christian Schneider,$^{1,}$}

\email
{Corresponding author: \\christian.schneider@uni-oldenburg.de}

\affiliation
{
$^{1}$Institute of Physics, Faculty V, Carl von Ossietzky University Oldenburg, 26129 Oldenburg, Germany\\ 
$^{2}$Department of Physics and Astronomy, University of Exeter, Exeter EX4 4QL, United Kingdom\\
$^{3}$Department of Physics, Xiamen University Malaysia, 49300 Sepang, Malaysia\\
$^{4}$School of Physics and Optoelectronic Engineering, Beijing University of Technology, Beijing 100124, China\\
$^{5}$Technion-Israel Institute of Technology, 32000 Haifa, Israel\\
$^{6}$Physics Program, Guangdong Technion-Israel Institute of Technology, Shantou, Guangdong 515063, China\\
$^{7}$Guangdong Provincial Key Laboratory of Materials and Technologies for Energy Conversion, Guangdong Technion-Israel Institute of Technology, Shantou, Guangdong 515063, China\\
$^{8}$Institute of Applied Physics, Abbe Center of Photonics, Friedrich Schiller University Jena, 07745 Jena, Germany\\
$^{9}$Fraunhofer-Institute for Applied Optics and Precision Engineering IOF, 07745 Jena, Germany\\
$^{10}$Max Planck School of Photonics, 07745 Jena, Germany\\
$^{11}$Department of Inorganic Chemistry, Faculty of Chemical Technology, University of Chemistry and Technology Prague, Technick\'{a} 5, Prague 6, 16628, Czech Republic\\
$^{12}$School of Mathematical and Physical Sciences, University of Sheffield, Sheffield S10 2TN, United Kingdom\\
$^{13}$These authors contributed equally to this work. *Corresponding author: christian.schneider@uni-oldenburg.de
}

\date{\today}

\maketitle

\textbf{Van der Waals magnets are an emergent material class of paramount interest for fundamental studies in coupling light with matter excitations, which are uniquely linked to their underlying magnetic properties. Among these materials, the semiconducting magnet CrSBr is possibly a first playground where we can study simultaneously the interaction of photons, magnons, and excitons at the quantum level. Here we demonstrate a coherent macroscopic quantum phase, the bosonic condensation of exciton-polaritons, emerging in a CrSBr flake embedded in a fully tunable cryogenic open optical cavity. The Bose condensate is characterized by a highly non-linear threshold-like behavior, macroscopic occupation of the ground state and coherence manifests distinctly via its first and second order quantum correlations. We find that the condensate's non-linearity is highly susceptible to the magnetic order in CrSBr. Specially, it can encounter a sign change from attractive to repulsive interactions when the intrinsic antiferromagnetic order transforms to the forced ferromagnetic order. Our findings open a route towards magnetically controllable quantum fluids of light, and optomagnonic devices where spin magnetism is coupled to on-chip Bose-Einstein condensates.}

\section{Introduction} 

\begin{figure*}[t]
\includegraphics[width=2\columnwidth]{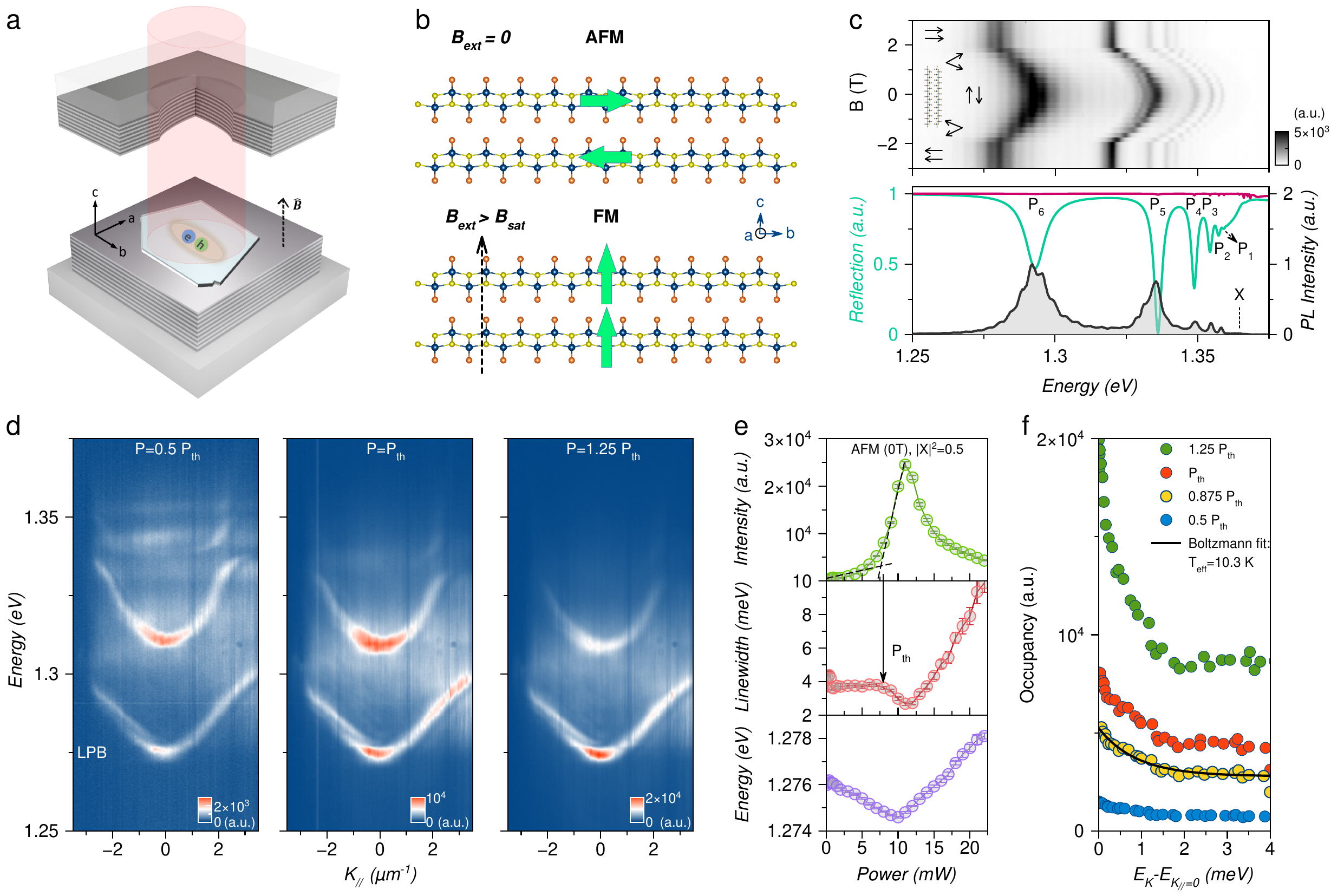}
    \caption{\textbf{Exciton-polariton condensation of CrSBr in a 2D planar microcavity.} \textbf{a.} Schematics of the cryogenic open optical microcavity. Excitonic optical dipoles align to the in-plane crystallographic b-axis. \textbf{b.} Schematics of the magnetic orders in CrSBr with Cr, S and Br atoms in blue, yellow and red colors, respectively. The green arrows indicate the intralayer net magnetization. Top panel: intrinsic anti-ferromagnetic (AFM) ground state below Néel temperature. Bottom panel: forced ferromagnetic (FM) order in an out-of-plane saturation field B$_{\mathrm{sat}}\sim$2 T. \textbf{c.} bottom panel: PL measurement (black), and reflectivity (green) of a 312 nm thick CrSBr flake on bottom DBR simulated by transfer matrix method with (green) and without (magenta) excitonic resonance (1.3655 eV). Six self-hybridized polariton states (P$_1$-P$_6$) are observed. Top panel: magneto-PL of P$_1$-P$_6$ in the bare flake. Spin canting is illustrated for a bilayer case.  \textbf{d.} Polariton condensation in the momentum-space PL measurements. Macroscopic occupation of the polariton ground state occurs at elevated pump powers. \textbf{e.} Top to bottom: Integrated emission intensity, linewidth and energy as a function of excitation power of the ground state at $K_{//}$=0. \textbf{f.} Polariton occupancy in ground and excited polariton states for excitation powers across the threshold (P$_{th}$).}
    \label{figure_1}
\end{figure*}

The emerging class of van der Waals magnets offers unprecedented opportunities to interface magnetism with matter excitations at the atomic limit \cite{gibertini2019magnetic}. Among these materials, CrSBr is of specific interest due to its ambient stability and semiconducting properties \cite{ziebel2024crsbr}. The excitons in CrSBr are prominent down to the monolayer limit \cite{wilson2021interlayer,tabataba2024doping}, and dictate the optical response over a large temperature range \cite{lin2024strong,dirnberger2023magneto,linhart2023optical}. From bulk to few layers, the magnetic ground state is represented by an A-type interlayer antiferromagnetic (AFM) order below the Néel temperature $\sim$132 K \cite{lee2021magnetic,wilson2021interlayer,rizzo2022visualizing}, as shown in Fig. \ref{figure_1}b. In AFM phase, excitons are strongly confined within the individual layer due to the spin-forbidden interlayer charge transfer \cite{wilson2021interlayer,heissenbuttel2025quadratic,shao2025magnetically}.

The magnetic order can be controlled by strain and hydrostatic pressure \cite{cenker2022reversible,pawbake2023magneto,cenker2025strain}, electrostatic doping \cite{tabataba2024doping,hong2025charge}, externally applied magnetic fields\cite{wilson2021interlayer}, and recently demonstrated twist-engineering \cite{mondal2025twist}. Out-of-plane magnetic fields can force the system into a parallel spin configuration of staggered ferromagnetic (FM) order (Fig. \ref{figure_1}b). The exciton energy changes drastically with the interlayer hybridization \cite{wilson2021interlayer,dirnberger2023magneto}, and thus represents a direct optical read-out channel for the magnetic order \cite{sun2025resolving}. In presence of external magnetic fields, the excitonic landscape is further modulated by magnons, which emerge in the system and have recently been observed in pump-probe studies \cite{bae2022exciton,dirnberger2023magneto,komar2024colossal,diederich2023tunable,datta2025magnon}. These phenomena attest strong interplay between the magnetic-electronic-optical properties in CrSBr. The interplay between magnetic order and the light-matter coupling in optical cavities, to date, has been explored less intensely. First reports have verified the emergence of strongly coupled exciton-polaritons in bare CrSBr slabs \cite{dirnberger2023magneto} and those embedded in closed optical cavities \cite{dirnberger2023magneto,wang2023magnetically,li2024two,ruta2023hyperbolic}, and highlighted the magnetic response of the hybrid light-matter quasi-particles. However, these studies were performed in the linear regime, where exciton correlations can be widely neglected in the dilute polariton gas. 

Since the polariton mass can reach values comparable to the cavity photon mass \cite{weisbuch1992observation}, phase transitions into macroscopic coherent quantum states, including bosonic condensates, superfluidity or Kosterlitz-Thouless phases are more likely to occur at finite cryogenic temperatures. A major appeal thus arises to explore high-density regimes of exciton-polaritons in two-dimensional magnets, as coherent phases that interact with the underlying magnetic order remain elusive. 

Here, we use a cryogenic tunable optical microcavity with a high quality factor to acquire the necessary in-situ control of the light-matter composition of CrSBr exciton-polaritons for the transition to the non-linear regime of polariton condensation. The emergent exciton-polariton condensate is evidenced by the threshold-like emission output and linewidth reduction, a macroscopically occupied polariton ground state, and its distinct first and second order quantum coherence. The interplay of the condensate with the magnetic order furthermore yields an intriguingly new kind of tunable polariton non-linearity that is driven by the excitation of incoherent magnons. The exciton-polaritons in CrSBr show attractive and repulsive interactions in the AFM and FM orders, respectively.

\section{Optical properties}

For our study, we utilize a thin CrSBr flake, which we prepare via exfoliation and subsequent dry transfer on a SiO$_2$/TiO$_2$ distributed Bragg reflector (DBR). The thickness of the slab is determined as 312 $\pm$ 2 nm via atomic force microscopy (see Supplementary Fig. S1d). Owing to the drastic dielectric contrast between the CrSBr crystal and the surrounding medium of substrate materials and vacuum, the crystal by itself already composes a Fabry-Pérot cavity \cite{dirnberger2023magneto}, contrasting the situation of nanometer thin flakes or bulk crystals. The light-matter coupling strength of excitons in CrSBr was found to be sufficient to support the emergence of self-hybridized exciton-polaritons \cite{dirnberger2023magneto, wang2023magnetically}, while the energy of these intrinsic polaritonic resonances depend sensibly on the exact thickness of the CrSBr crystal. The optical transitions subject to our study thus need to be analyzed in a polaritonic picture. The photoluminescence (PL) spectrum of our flake at 3.5 K features six self-hybridized polariton peaks (P$_1$-P$_6$) in the bottom panel of Fig. \ref{figure_1}c, which are consistent with the reflection spectrum of a 312 nm thick CrSBr placed on a DBR, simulated by the transfer matrix method (see Methods and Supplementary Fig. S1). 

Exciton-polaritons are bosonic quasi-particles, whose magnetic response should arise from their excitonic component \cite{weisbuch1992observation,han2025infrared}. The out-of-plane magnetic field cants the electron spins away from the A-type AFM order in CrSBr, and the net magnetization of the crystal saturates as a forced FM order at $B_{\mathrm{sat}}\sim$ 2 T. As a result, the spin-allowed interlayer charge transfers take place \cite{heissenbuttel2025quadratic}, and the exciton wave-function expands into neighboring sheets. Thus, the energy of excitons and consequently the emergent self-hybridized polaritons encounter a spectral redshift as a result of the admixture of more spin-parallel bands \cite{wilson2021interlayer,heissenbuttel2025quadratic}. Further, direct evidence of the polaritonic nature of these peaks is reflected by their magnetic field response in the upper panel of Fig. \ref{figure_1}c: While all peaks sensibly react to an external out-of-plane magnetic field via a strong energy redshift, quantitatively, this shift depends on the peak energy. Indeed, the energetically lowest P$_6$ self-hybridized polariton features the smallest magnetic field shift, while the shift of the polariton modes at higher energies scales with their excitonic admixture that is characterized by the excitonic Hopfield coefficient \cite{hopfield1958theory,dirnberger2023magneto}. Specifically, -13.3 meV magnetic redshift of the intrinsic P$_6$ mode corresponds to an excitonic Hopfield coefficient: $|X|_{P_6}^2$= 0.76 (see Supplementary Note S2).    

Active control of the cavity resonances is acquired by introducing a highly reflective top DBR, as depicted in Fig. \ref{figure_1}a (see details in Supplementary Fig. S1). The top DBR features larger planar areas, as well as microstructured sections with hemispheric optical traps to provide in-plane mode confinement. The schematics capture the essence of the resulting tunable open-access cavity structure. In this setting, the gap between the top DBR and the CrSBr crystal is tunable via a DC voltage applied to the piezo nanopositioners. A maximum voltage of 60 V corresponds to a continuous travel range $\sim$0.6 $\mu$m.

First, we study the resulting coupled system in the planar area of the open cavity via recording momentum-resolved PL emission spectra. The system is excited with off-resonant 725 nm laser pulses with a temporal pulse duration of 200 fs and a repetition rate of 76 MHz (see Methods). The laser polarization is aligned to the crystal b-axis, and the laser beam is focused down to a spot of 6 $\mu$m in diameter. The size of the air-gap keeps constant at a value of 3.31 $\mu$m to establish the hybridization of the external cavity resonance with the lowest self-hybridized P6 polariton mode. This is unambiguously verified in the left panel of Fig. \ref{figure_1}d, via the avoided crossing features around all of the self-hybridized polariton modes at finite in-plane wavevectors (K$_{//}$). We note that the effective Hopfield coefficient of the lowest polariton branch (LPB) resonance is |X|$^2$=0.5, and thus carries a substantial excitonic fraction (see Supplementary Note S2).

\section{Condensation of CrSBr exciton-polaritons in 2D planar and 0D discretized cavity modes} 

Being bosonic quasi-particles, exciton-polaritons can undergo a phase transition into a coherent state at elevated density, while maintaining the inherent non-linear character of the excitons and retaining the fingerprints of specific magnetic order. As an exemplary demonstration of the formation of a coherent macroscopic quantum state, we tentatively ramp up the excitation power, and witness a qualitative modification of the quasi-particle occupation, featuring a relaxation to the ground state of the lowest polariton branch, while higher energy and higher momentum modes encounter a depletion of quasi-particles (see Fig. \ref{figure_1}d). In Fig. \ref{figure_1}e, we plot the Lorentzian fit results including the emission intensity, linewidth and energy of the ground state as a function of the excitation power. With increasing excitation power, the emission intensity encounters a clear non-linearity with a distinct threshold-like behavior $\sim$8 mW. Simultaneously, the linewidth encounters a significant narrowing by 1.1 meV before a strong line broadening emerges due to the dissipative effects. The energy of the polariton ground state encounters power-dependent energy shifts on the meV scale, two orders of magnitude smaller than the Rabi splitting of CrSBr polaritons \cite{smolenski2025large}, also featuring a rather unique crossover from a moderate redshifting to a blueshifting behavior. As we will detail in next section, this non-linear energy shift of the LPB emerges as a consequence from interactions between polaritons and incoherent magnons.  

We plot the extracted quasi-particle occupancy in Fig. \ref{figure_1}f (see Methods). Around the threshold, the distribution of the polariton gas follows a Boltzmann distribution with a temperature of T$_{eff}$=10.3 K for P=0.875 P$_{th}$. Above threshold at P=1.25 P$_{th}$, the ground state clearly becomes macroscopically occupied and the distribution markedly deviates from a Boltzmann distribution.

\begin{figure*}[t]
\includegraphics[width=2\columnwidth]{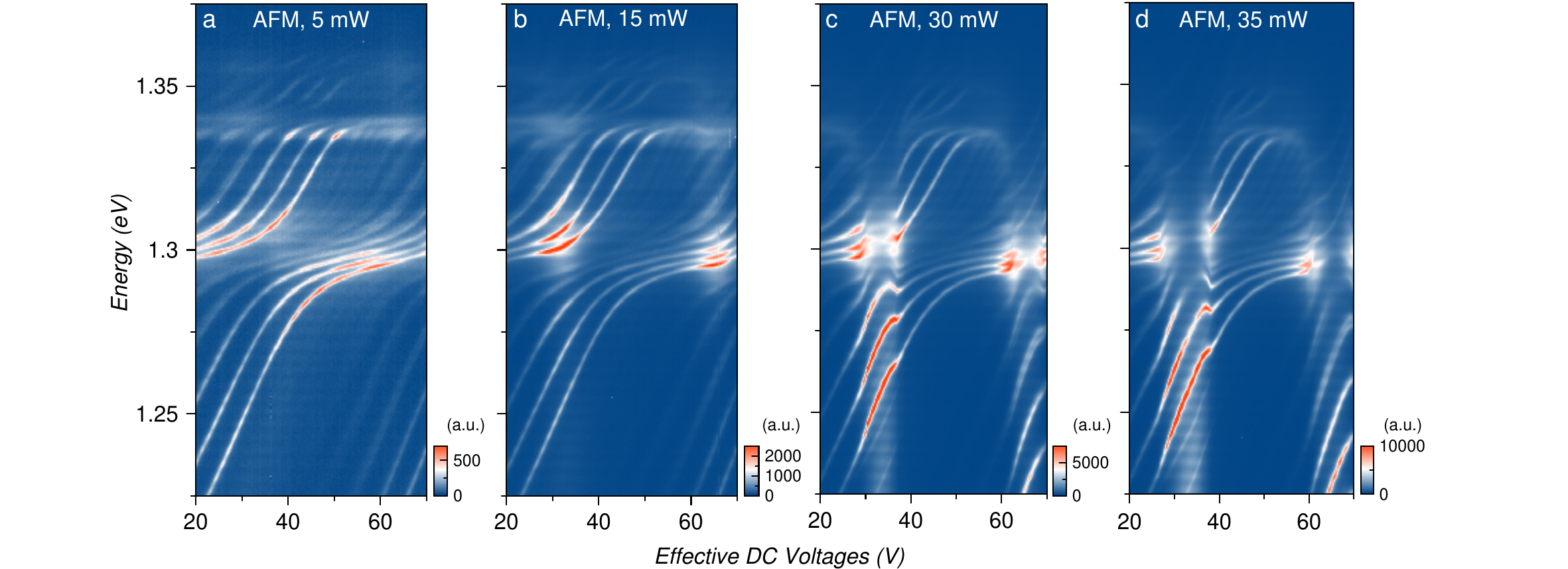}
   \caption{\textbf{Stimulated polaritonic scattering in 0D cavity.} The threshold phenomena is clear around \textcolor{red}{15} mW pump power in \textbf{b}. Note that the cavity length is slightly reduced relative to the scenarios in Fig. \ref{figure_3}a, to monitor the stimulated scattering in two consecutive longitudinal mode orders. An effective DC voltages above 60 V are thus used to characterize the extra detunings. The condensation phenomena into the LPBs with effective DC$\sim$32 V are more prominent for high pump densities.}
\label{figure_2}
\end{figure*}

To have a more accurate probe of the polariton energy shift and precise control of the cavity detuning energy, we resort to the discretized zero-dimensional (0D) cavity modes by using the concave-shaped DBR in the top mesa. The concave-shaped DBR can confine Laguerre-Gaussian modes of up to five varying transversal mode orders within a longitudinal mode order. These discretized 0D modes \cite{dusel2020room} have complex spatial profiles in real-space or momentum-space that have been explored via tomographic measurements \cite{horneber2024enwrapped,dufferwiel2014strong}, and their theoretical quality factors (Q-factors) are estimated to exceed 5000 and that for the 2D planar cavity (see Methods and Supplementary Fig. S1). 0D polaritons have meanwhile been established as a system of enhanced purity, in relation to polariton formation \cite{dufferwiel2014strong,dufferwiel2015exciton,sidler2017fermi,emmanuele2020highly,han2025infrared}, and specifically for condensation as well as polariton lasing experiments \cite{bajoni2008polariton,kim2016coherent}. In our case, the energetically lowest transverse mode has a Gaussian-shaped profile, matching the Gaussian laser excitation beam used in our measurements, so that the PL intensity emitted through this mode is maximized and we will only focus on the lowest transversal cavity modes in this work.

The applied maximum fluence energy is up to 0.66 nJ per pulse, corresponding to a time-averaged pump power of 50 mW. A concave top DBR of 6 $\mu$m in diameter is then utilized to match the focused laser spot. We use effective DC voltages to denote additional detunings (DC>60 V) that are beyond the default detuning range. As we can see from Fig. \ref{figure_2}a, the various sets of anti-crossings for each transverse cavity mode develop around the self-hybridized polaritons as we spectrally tune the discretized open-cavity resonances. This peculiar behavior is a clear signature of new exciton-polariton states, which hybridize the external high-Q 0D modes with the self-hybridized polaritons in the CrSBr slab. Furthermore, in the pump power dependent cavity detuning PL in Fig. \ref{figure_2}, we additionally observe unambiguous and strong scattering phenomena into the LPBs for DC=27-37 V ($|X|^2$=6-27 \%), which form an essential relaxation pathway in forming a polaritonic condensate. Fig. S3a shows the experimental results of a more detailed power dependent PL for DC=32 V, where the scattering is the most pronounced. The corresponding energy difference between the $P_6$ self-hybridized polariton and the LPB is 41 $\pm$ 6 meV, where the uncertainty accounts for the linewidth of $P_6$ obtained from Fig. \ref{figure_1}c. This energy difference matches with the $A^3_{g}$ optical phonon $\sim$43 meV in CrSBr, which corresponds to lattice vibrations primarily involving anti-phase vibration of Chromium (Cr) and Sulfur (S) atoms \cite{mondal2025raman,pawbake2023raman}. This remarks efficient phonon-assisted polariton relaxation in the condensation process. 


\section{Magnetic-phase dependent polaritonic non-linearity.}

\begin{figure*}[t]
\includegraphics[width=2\columnwidth]{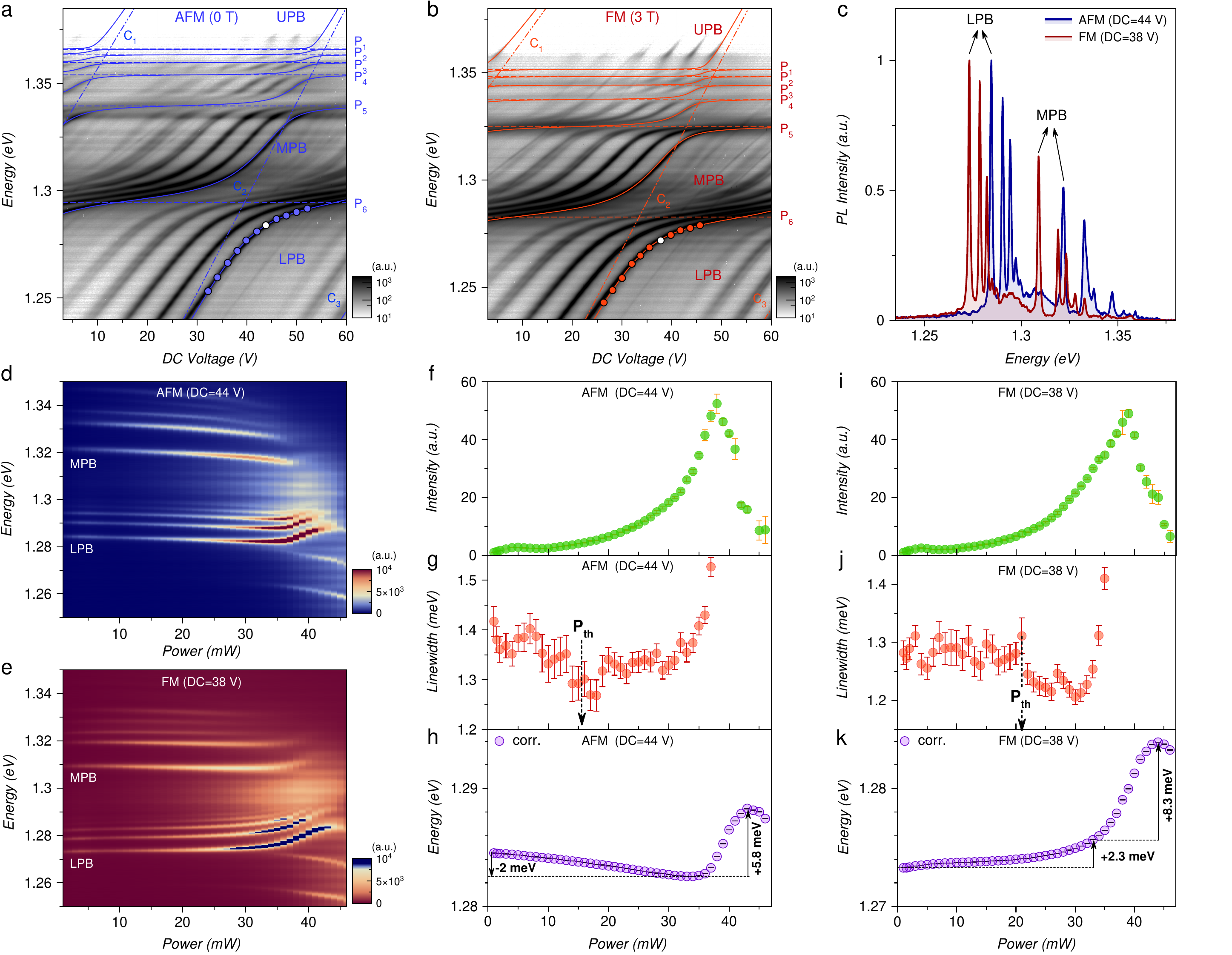}
    \caption{\textbf{Pump power dependent PL measurements of the polariton modes using discretized cavity modes.} Low-power PL measurements with different cavity detunings (DC voltages) in \textbf{a.} AFM order at 0 T and \textbf{b.} FM order at 3 T. The superimposed lines are the fitting of the new polariton modes (solid), self-hybridized polaritons (dashed), and the cavity modes (dotted-dashed). The coupling strengths are summarized in Supplementary Table. I. The dots on LPBs mark the detunings for power dependent PL in Supplementary Fig. S3. \textbf{c.} PL spectra of the new polariton modes with same external cavity detuning marked by the white dots in \textbf{a} and \textbf{b}. Pump power dependent PL for \textbf{d.} AFM order (0 T, DC=44 V), and \textbf{e.} FM order (3 T, DC=38 V). \textbf{f.} Intensity, \textbf{g.} linewidth, and \textbf{h.} energy of the energetically lowest LPB mode in \textbf{d}. \textbf{i.} Intensity, \textbf{j.} linewidth, and \textbf{k.} energy of the energetically lowest LPB mode in \textbf{e}. All data are fitted via a Lorentzian function. The threshold is around P$_{th}$=16 mW (21 mW) for AFM (FM) order (see logarithm scale plot also in Supplementary Figs. S4i and S4j). The error bars of the energy shifts in \textbf{e} and \textbf{h} are smaller than the symbol size. The experimentally measured polariton energy shifts that are the hollow dots in Figs. S4a and S4b are already corrected in (e, h) by removing the redshift due to cavity thermal expansion with increasing pump power (see Supplementary Note S2).}
    \label{figure_3}
\end{figure*}

Having established the condensation process and ground-state scattering of exciton-polaritons in our system, in the following, we investigate the influence of magnetic order on the condensate's non-linearity via density dependent PL measurements. Figures. \ref{figure_3}a and \ref{figure_3}b present the low-density PL spectra as a function of the cavity detuning voltage in the AFM order at 0 T and FM order at 3 T, respectively (see Methods). The PL emission from the tunable cavity polaritonic modes is dominated by the lower polariton branch (LPB), five middle polariton branches (MPB) that have reduced intensity and weak spectral features of the upper polariton branch (UPB). We denote the UPB with respect to the external coupling to P$_1$ mode in our special case, but not those due to self-hybridization.

This strong coupling cascade can be fully modeled consistently via the transfer matrix method, by adding the tunable top DBR and the variable cavity gap on top of the structure used in Fig. \ref{figure_1}c (see Supplementary Fig. S1). While we obtain excellent agreement between model and experiment, we furthermore resort to a more intuitive description in the framework of an extended coupled oscillator model, which is more straightforward for further calculating the Hopfield coefficients in presence of the external cavity structure. Therein, we consider six independent self-hybridized polariton states (P$_1$-P$_6$) and the Gaussian transversal modes (C$_1$-C$_3$) from three consecutive longitudinal mode orders (see Methods). More than one cavity mode participate coherently in the strong coupling and consequently bend the new polariton energies even above (and below) the P$_6$ self-hybridized polariton state above 55 V (and below 15 V). As a result, the dispersion of the energetically lowest MPB connects directly to the LPB via the transversal modes of the same symmetry but different longitudinal mode orders. 

This phenomenon has been observed more clearly from the power-dependent PL measurements of the cavity scan in Fig. \ref{figure_2}. Diagonalizing the 9$\times$9 matrix \ref{coupled_oscillator} gives us the eigenvalues of nine new exciton-polariton modes at different cavity detuning voltages. The maximum coupling strength of V$_6$= 33 meV corresponds to the strong coupling of the cavity modes to P$_6$. This result is in contrast to the previously reported ultra-strong coupling regime of CrSBr excitons in a closed cavity structure \cite{wang2023magnetically}, since our polaritons arise from the resonant coupling of the self-hybridized polaritons with the external open cavity resonances. 

Similar to the spectral shift of the intrinsic self-hybridized polaritons, a magnetic response occurs for all of the new polariton modes in our open cavity (see Supplementary Figs. S2c-S2r). Notably, the amplitude of the magneto-polariton energy shifts depends sensibly on the in-situ adjustable detuning, which manifest the explicit Hopfield physics of the exciton-polaritons. This is most instructively reflected by the LPB that remains almost fully insensitive to the external B-field at detuning of DC=20 V (higher photonic regime in Fig. S2r), and develops a considerable magnetic field shift at DC=50 V (higher excitonic regime in Fig. S2c) that is comparable to that of the self-hybridized P$_6$ mode in Fig. S2b. 

Fig. \ref{figure_3}b presents the PL spectra as a function of cavity detuning in a fixed saturation magnetic field of 3 T. In contrast to the AFM polaritons in Fig. \ref{figure_3}a, the FM polaritonic resonances are collectively redshifted due to the layer hybridization effect. We notice that the coupling strengths of the self-hybridized polaritons are nearly invariant to the applied magnetic field (see Supplementary Note S1, Table. S1), consistent with recent magneto-optics measurements of a much thinner 7-layer CrSBr placed on a photonic crystal cavity \cite{li2024two}. The Hopfield coefficients of the P$_6$ self-hybridized polaritons in our external cavity ($|P_6|^2$ and ($|C_2|^2$) can be found in Supplementary Figs. S4e and S4f. For the AFM and FM orders, a same detuning condition relative to the P$_6$ self-hybridized polaritons has a DC voltage difference of 6 V. 

Two exemplary PL spectra with the same cavity detuning condition that are marked by the white dots in Figs. \ref{figure_3}a and \ref{figure_3}b are plotted in Fig. \ref{figure_3}c, where the excitonic Hopfield coefficients of the LPBs are $|X|^2$=$|X|_{P_6}^2\times|P_6|^2$= 0.56 (see calculation details in Supplementary Note S2). Same detuning conditions as in Fig. \ref{figure_3}c are thus chosen to warrant comparability between the magnetic-order contrasting polaritonic non-linearity. Figures \ref{figure_3}d and \ref{figure_3}e depict the power dependent evolution of the discrete modes in the LPBs and the MPBs. Both sets of measurements feature qualitative similarities: the LPB modes of lowest energy progressively gain intensity, and eventually experience a pronounced blue-shift at pump powers above 33 mW, whereas the MPBs deplete in population in the high density regime, which clearly reflects the Bosonic final state stimulation \cite{dang1998stimulation}, as has been observed in Fig. \ref{figure_1}d and Fig. \ref{figure_2}. 

Similar to the 2D planar results in Fig. \ref{figure_1}e, both 0D experimental data recorded at 0 T and 3 T reveal a pronounced superlinear increase in emission intensity over a large power range of 15-35 mW (see Figs. \ref{figure_3}f and \ref{figure_3}i). As shown in Figs. \ref{figure_3}g and \ref{figure_3}j, this pronounced non-linear behavior is accompanied by a reduction of the emission linewidth. We note that this lineshape can be broadened from fast vibrational fluctuations of the modes in the open cavity, probably due to the thermal effect on the open cavity (see discussion in Supplementary Notes S3, and top panel of Fig. S5b that shows substantial heat localization in top DBR). We also note that the polariton linewidth in our 0D discrete cavity is approximately 3 times smaller than that in the 2D planar cavity case, so that the moderate linewidth drop is reasonable, which is in principle in agreement with studies of 0D monolithic GaAs cavity with very high Q-factor\cite{bajoni2008polariton}. At largest pump powers above 33 mW, the full width at half maximum (FWHM) strongly increases, which we assign to density fluctuations in the LPB states, also in a good agreement with earlier reports on polariton condensates driven with ultrashort laser pulses \cite{kasprzak2006bose,tempel2012characterization,bajoni2008polariton}. Indeed, in this regime, the emission intensity in the ground state also experiences a drop, which further supports this picture \cite{tempel2012temperature}. 

For the maximum pump power P=50 mW, we estimate the exciton density to be $d_X^{50 mW}$= 4.54$\times 10^{11}$ $cm^{-2}$ in each layer of our CrSBr flake (see details in Supplementary Note S3 and Fig. S5c), which is well below the Mott density $n_M\sim$$10^{14}$ $cm^{-2}$ estimated for the CrSBr in a recent study \cite{dirnberger2026exciton} or even the $n_M\sim$$10^{13}$ $cm^{-2}$ in a typical semiconducting transition metal dichalcogenide monolayer \cite{chernikov2015population,wang2018colloquium}. Considering the effective Bohr radius is much smaller in CrSBr \cite{klein2023bulk,semina2024excitons,li2024two,smolenski2025large}, we expect the Mott density in a CrSBr layer to be at least one order of magnitude higher than $d_X^{50 mW}$. Therefore, we suppose our system has not reached the Mott transition even for the maximum applied pump power (detailed discussion in  Supplementary Note S3). 

For both magnetic orders in Figs. \ref{figure_3}h and \ref{figure_3}k, the optical resonance displays two distinct regimes: Up to pump powers around 33 mW, the spectral shifts develop linearly with the pump power, and have modest values of a few meV. In addition, we capture a super-imposed redshift feature of the LPB for P<20 mW in the experiments, which we attribute to a slight thermal expansion of our open cavity (see Supplementary Note S2 and Fig. S5b top panel), in combination with the interaction between the excitons and incoherent magnons. We can accurately account for the artificial redshift of the cavity modes due to thermal expansion, by referencing the purely photonic resonances (see details in Supplementary Note S2). 

Figures \ref{figure_3}h and \ref{figure_3}k show the resulting polaritonic energy shifts of the LPBs in AFM and FM orders. In the AFM order at 0 T, the LPB redshifts by -2 meV until P=33 mW, while it blueshifts with a similar amplitude of +2.3 meV in the FM order at 3 T. Such opposite trends are general for all of the LPBs with same magnetic orders. We have also performed the detailed pump power dependent measurements for various cavity detunings that are marked by the color dots in Figs. \ref{figure_3}a and \ref{figure_3}b, whose results are shown in Supplementary Fig. S3. That is to say, in the case of AFM order, the LPB modes encounter a net red-shift for all detuning conditions, while the LPB modes in the FM order experience a contrasting blueshift for all detunings (see Supplementary Figs. S4c and S4d). These trends become also more prominent with increasing excitonic coefficient (Supplementary Fig. S4g solid symbols), which is an explicit evidence on the magnetic-order and density dependent excitonic interactions mediated by the incoherent magnons \cite{dirnberger2023magneto}. 


To gain further insights into the nature of non-linear response, we theoretically investigate the microscopic origin for different non-linear processes that lead to the observed energy shift (details in Supplementary Note S4). As the leading contribution in the regime of low to modest exciton densities, we identify the combined effect of exciton-exciton interactions and the influence of incoherent magnon excitations. Note that here as a magnon mode we consider spin waves at $k=0$ (Kittel mode corresponding to the uniform precession), since coupling to distinct modes with non-zero momentum is suppressed due to selection rules and significant exciton broadening. The corresponding energy shift depends on the magnon mode occupation, and can be parametrized by the effective temperature. The shift is defined as $\Delta\mathcal{E}\approx\mathcal{A}(B) \rho_\mathrm{X} + \mathcal{B}(B)\Delta T$, where $\rho_\mathrm{X}$ is the exciton density and $\Delta T$ is the change of magnon temperature. 

The first term in $\Delta\mathcal{E}$, proportional to the $\mathcal{A}$-factor, comes from Coulomb-based exciton-exciton (X-X) interactions \cite{Shahnazaryan2017} and non-linear phase space filling effects \cite{Song:PRR6-2024}. Both contributions lead to the blueshift of LPB energy at increasing pump power (exciton occupation $\rho_{\mathrm{X}}$). This blueshift is of non-magnetic origin and is present for both zero and finite magnetic fields. Studying the excitonic properties in CrSBr, we estimate the corresponding interaction  $\mathcal{A}(B)=0.48~\mu$eV$\mu$m$^2$, and find that it stays approximately constant in the magnetic fields from $B=0$ to $B=3$~T. 

The second term in $\Delta\mathcal{E}$ arises from the coupling between excitonic and magnonic modes, being a signature of the magnetic property of CrSBr. Here, we need to consider cases of AFM and FM ordering separately. We find that while magnon-based contributions to the non-linearity vanish in zero and saturating magnetic fields, the presence of incoherent magnons however can introduce a strong redshift for the LPB mode in the AFM case. This corresponds to the temperature-dependent term with the $\mathcal{B}$-factor, which depends strongly on the underlying magnetic order (see Supplementary Note S4). Analyzing the magnonic spectrum, we find that in AFM order the $\mathcal{B}$-factor is significant and can be approximated as $\mathcal{B}(0)$=$-32k_B$, where $k_B$ is the Boltzmann constant. The overall behaviour corresponds to the redshift or attractive interaction, reflecting the experimental observations in Fig. \ref{figure_3}h and Supplementary Fig. S4c. In the FM case, the magnonic contribution is very weak ($\mathcal{B}(\text{3T})\approx 0$), resulting in a net blueshift that is predominant due to the X-X interaction and phase space filling effects, described by the $\mathcal{A}$-factor, and corresponding to the repulsive interactions of the polaritons.


The non-linear features of our polaritons become most pronounced in the power-dependent blueshift for pump powers exceeding $33$~mW in Figs. \ref{figure_3}h and \ref{figure_3}k. The non-linear shift for the LPB in both magnetic orders reaches highest values around $44$~mW, before it saturates at even higher pump powers. Starting from the turning point of redshift-to-blueshift of the LPB in AFM order, the maximum blueshift amplitude in FM order is +$8.3$~meV, significantly larger than the +$5.8$~meV blueshift in AFM order. The strong increase of the blueshift in the high density regime can no longer be assigned to X-X interactions but aligns with the saturation of the Rabi-splitting via phase space filling (PSF), which results in a significant blueshift of the P$_6$ resonance and consequently translates to a similar energy shift of the hybridized open cavity mode. This magnetically controllable non-linear interaction in CrSBr is really unconventional. We note that this saturation behavior, which displays a clear dependency on magnetic order, hints at the interplay between saturation and interlayer localization-delocalization of the exciton wavefunction via spin-allowed layer coupling. According to Eqs.~S(56) and S(57) in Supplementary Note S4, we predict the change from AFM order to FM order leads to at least 20 \% enhancement in the optical saturation effect.

\begin{figure*}[t]
\includegraphics[width=2\columnwidth]{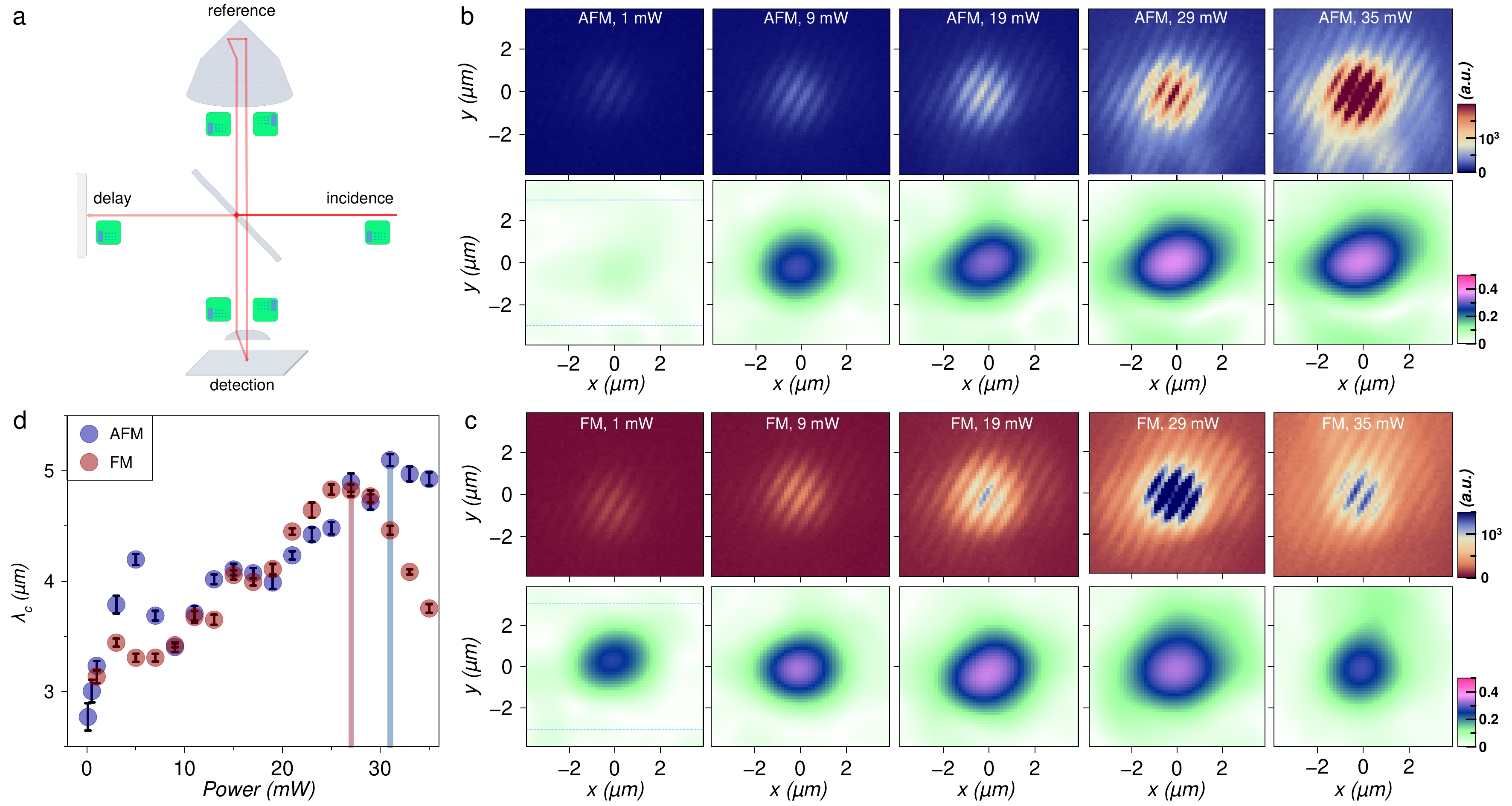}
    \caption{\textbf{Magnetic order dependent first-order correlation measurements of the exciton-polariton condensate.} \textbf{a.} Schematics of the Michelson interferometer. Spatial coherence measurements of the LPB in \textbf{b.} AFM order and \textbf{c.} FM order (low-density spectra as in Fig. \ref{figure_3}c). Upper panels: power dependent spatial interference at zero-delay ($\Delta\tau=0$). Lower panels: calculated spatially-resolved first-order correlation function g$^{(1)}(\vec{r},0)$ at pump powers corresponding to the upper panels. The two dashed lines in the two bottom left graphs show a 6 $\mu$m confined region for vertical binning of g$^{(1)}(\vec{r},0)$, which is then fitted by a Gaussian function to extract the FWHM 'x$_c$' in the horizontal direction. \textbf{d.} Magnetic order and power dependent coherence length $\lambda_c$=$\sqrt{\pi}x_c$. Two vertical lines indicate the pump powers for the onsets of coherence degradation.}
    \label{figure_4}
\end{figure*}

To be more quantitative, we can acquire the PSF blueshift ratio between the FM and AFM orders that have the same cavity detuning conditions. We consider the detuning conditions of DC=32-44 V in Fig. \ref{figure_3}a for AFM order and DC=26-38 V in Fig. \ref{figure_3}b for FM order. The corresponding excitonic Hopfield coefficients are between 12-56 \% (see calculative method in Supplementary Note S2). The power range of their blueshift regions are marked by the shades in Supplementary Figs. S4c and S4d. The blueshift ratio is then summarized in Supplementary Fig. S4h. We can see that the non-linear phase space filling is indeed more pronounced for the FM order due to the larger exciton wavefunction as the spin-allowed layer tunneling is switched on by the external magnetic field. 


\section{First-order spatio-temporal correlation}
Polariton condensates are a coherent quantum phase \cite{snoke2002spontaneous}, displaying specific fingerprints in the first-order \cite{kasprzak2006bose,deng2007spatial} and second-order correlation function \cite{deng2002condensation,kasprzak2008second,love2008intrinsic,fischer2014spatial}, especially at the transition from the linear- to the non-linear regime. In the following, we keep the magnetic field geometry and the cavity detuning conditions for the AFM and FM orders as in Figs. \ref{figure_3}d and \ref{figure_3}e.

First, we probe the emergence of first-order spatio-temporal coherence as a function of the polariton density. The experiment is carried out via a Michelson interferometer with a retro-reflector at the end of the reference arm, as sketched in Fig. \ref{figure_4}a. The resulting spatially inverted and slightly laterally shifted image is brought to overlap with the delay arm image by a focal lens (see Methods). A time-delay can be introduced via tuning the delay arm. The emergent interference fringes, which evolve with the polariton density, can be used to deduce the first-order spatio-temporal coherence: 
\begin{equation}
g^{(1)}(\vec{r},\Delta\tau)=\dfrac{I-I_1-I_2}{2\cos(\Delta\phi)\sqrt{I_1I_2}}
\label{1st_order_correlation}
\end{equation}
where $I$, $I_1$ and $I_2$ are the intensities of the interference, delay and reference arms, respectively. The phase difference $\Delta\phi$=$\Delta\vec{k}\cdot\vec{r}+\omega\Delta\tau$ is a function of the real-space position $\vec{r}$ of the interference fringes, wavevector difference $\Delta{\vec{k}}$ of the emissions from two arms, frequency $\omega$ of the condensate and interferometer time delay $\Delta\tau$. We see a clear development of the characteristic spatial interference fringes with the polariton density for the AFM and FM orders in the upper panels of Fig. \ref{figure_4}b and Fig. \ref{figure_4}c, respectively, which translates to the emergence of first-order correlation that clearly enhances with the excitation power (lower panels). Details on the data analysis are given in Methods.

To have a better notion of the spatial extension of the coherence, we summarize the power dependence of the coherence length $\lambda_c$ in Fig. \ref{figure_4}d. The condensate in AFM order reaches a maximum coherence length $\lambda_c$= 5.1 $\mu$m at a pump power of $\sim$30 mW, while the condensate in FM order reaches its maximum $\lambda_c$= 4.8 $\mu$m much earlier at $\sim$25 mW. This result is consistent with the larger polaritonic non-linearity owing to the expanded exciton wavefunction in the FM order, and once again pinpoints the impact of the magnetic ordering on the emergent macroscopic coherence in the quantum fluid. 

We assume that our system is not in thermal equilibrium at low pumping conditions. Note that full equilibrium has only been reported in very recent works on ultra-high Q-factor GaAs cavities \cite{sun2017bose}. For typical polariton systems studied under comparable conditions (meaning similar loss rates reflected by polaritonic linewidth $\sim$ 1 meV), temperature values in the range between 20 K \cite{kasprzak2006bose} and as large as 100 K \cite{balili2007bose} were reported. Therefore, going beyond T$_{eff}$=10.3 K that is acquired for the 2D planar cavity in Fig. \ref{figure_1}e, a reasonable estimate of the temperature of our 0D polariton gas on the order of 30 K yields a thermal \textit{de Broglie wavelength}: $\lambda_{dB}=(\dfrac{2\pi\hbar^2}{m_{\mathrm{eff}}^{LPB}k_BT})^{1/2}\approx$ 3 µm, where $m_{eff}^{LPB}$ is the effective mass of the LPB (see Supplementary Fig. S1g). $\lambda_{dB}$ indeed approaches our low-density measurement of the spatial coherence length $\lambda_c$ in Fig. \ref{figure_4}d. As pump power increases, $\lambda_c$ clearly exceed $\lambda_{dB}$, and reach the maxima of $\sim$5 $\mu m$ for both magnetic orders, which again in consistent with the onset of polariton condensation and the spatial confinement size of the 6 $\mu m$ concave DBR.


\section{Second-order temporal correlation}
A quantum optical characterization of the emergence of coherence in the light-matter coupled system is conducted via measuring the second-order temporal correlation of the condensates. The second-order correlation function reads
\begin{equation}
g^{(2)}(\tau)=\dfrac{\braket{I(t)I(t+\tau)}}{\braket{I(t)}\braket{I(t+\tau)}},
\label{2nd_order_correlation}
\end{equation}
where \textit{I($t$)} and \textit{I($t+\tau$)} are the emission intensities that are proportional to the number of photon counts registered at time $t$ and a delayed time $t+\tau$, respectively. The bracket denotes the time average. $g^{(2)}(\tau)$ characterizes the intensity correlations and thus gives insights into the statistics of photon emission from our system. We are particularly interested in the auto-correlation value at zero delay ${g^{(2)}(0)}$, which distinguishes a classical (${g^{(2)}(0)}$>1) from a coherent (${g^{(2)}(0)}$=1) and sub-Poissonian (${g^{(2)}(0)}$<1) emission statistics \cite{loudon2000quantum}. To highlight that the chosen experimental conditions yield an effective time averaging effect, which depends on the effective polariton relaxation time, we put an average sign on $g^{(2)}(0)$ to denote the averaging over the polariton lifetime upon pulsed excitation.

\begin{figure}[t]
\includegraphics[width=\columnwidth]{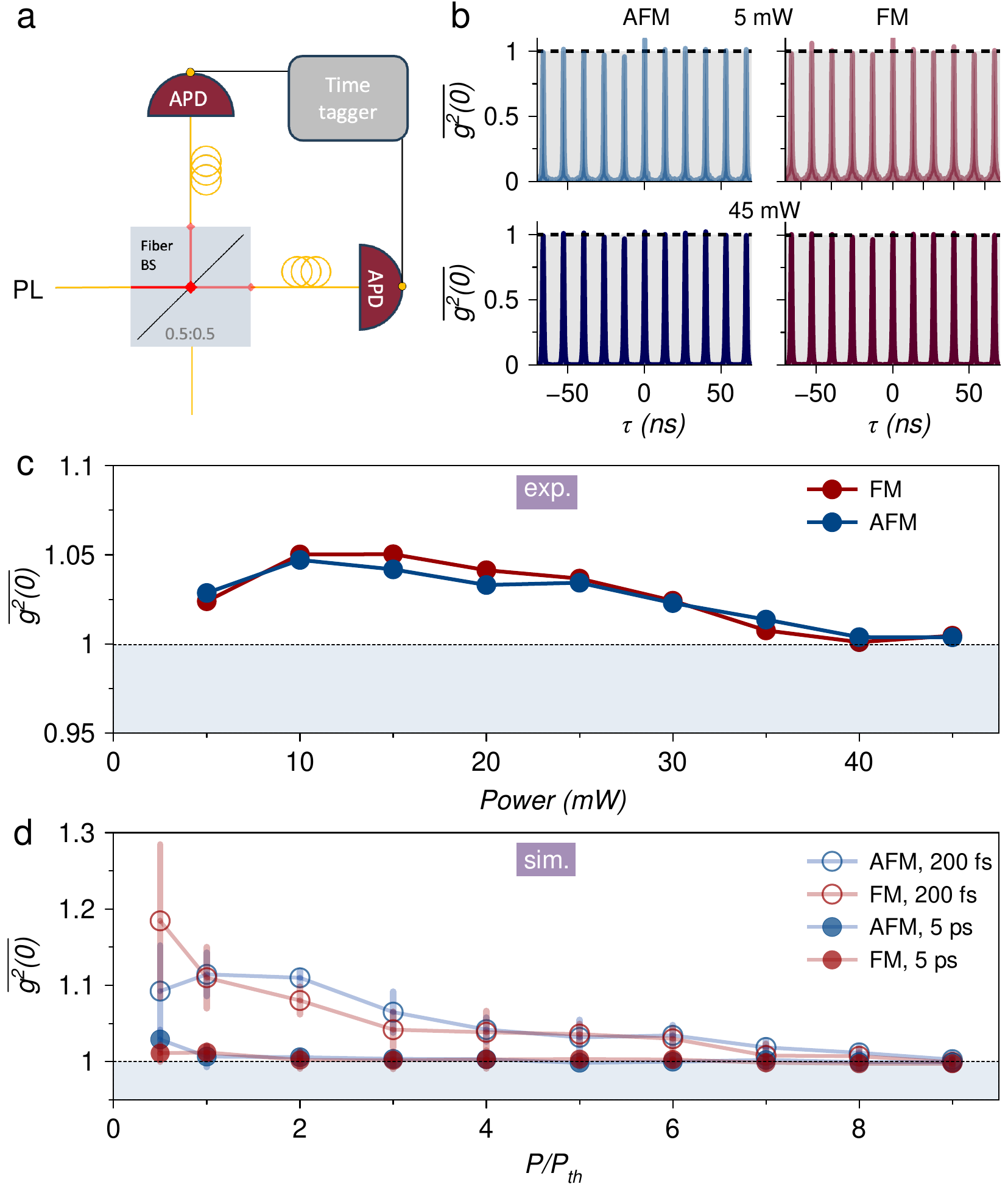}
    \caption{\textbf{Magnetic order dependent second-order correlation measurements and simulations of the exciton-polariton condensate.} \textbf{a.} Schematics of the Hanbury Brown and Twiss setup. \textbf{b.} Representative second-order correlation function measurements ${g^2(\tau)}$ of the LPBs in Fig. \ref{figure_3}c. The flat dashed lines at the unity mark the Poissonian level.  \textbf{c.} Experimental values and \textbf{d.} numerical simulations of the power dependent second-order auto-correlation $\overline{g^2(0)}$ for AFM and FM orders. The simulation results are firstly averaged over independent 'numerical measurements' of $g^2(\tau)$ with corresponding standard derivation shown by the error bar. Typical averaged results of $g^2(\tau)$ are shown in Supplementary Figs. S7c-S7f. We then obtain $\overline{g^2(0)}$ by averaging $g^2(\tau)$ at the beginning of $200$ fs to account for the excitation pulse width as in experiments (empty symbols), and a trial exciton reservoir lifetime of 5 ps (solid symbols).}
    \label{figure_5}
\end{figure}

The experiment is carried out via a Hanbury Brown and Twiss setup \cite{brown1956correlation} (see the schematics in Fig. \ref{figure_5}a, and instrumental details in Methods). It is composed of two avalanche photo diodes connected via a timing electronic tagger whose temporal resolution is 350 ps. Therefore, the experiment temporarily averages fully over the emission bursts from the LPB excited by the cavity, which is driven with 200 fs laser pulses with a repetition rate of 76 MHz. Because of the short excitonic lifetime of approximately a few picoseconds in CrSBr \cite{meineke2024ultrafast}, the anticipated $\overline{g^2(0)}=2$ of the thermal state for the pump power far below the condensate threshold is thus averaged out by the instrumental temporal resolution \cite{shan2023second,klaas2018evolution,wiersig2009direct}. The main features of this experiment are compiled in Fig.~5b for the AFM and FM orders at two pump power extrema (see complete power dependent $g^2(\tau)$ measurements in Supplementary Fig. S8). The auto-correlation values $\overline{g^2(0)}$ are summarized in Fig. \ref{figure_5}c. We notice that $\overline{g^2(0)}$ in both magnetic phases show a  decreasing trend towards 1, as the pump power ramps up.

The monotonous decrease of the second order correlation function above the condensation threshold evolves from the open dissipative nature of our system, and is a fingerprint of persistent particle fluctuations even at elevated densities. It is in excellent agreement with prior reports on the second order coherence in confined GaAs cavity polaritons \cite{klaas2018evolution}, exciton-polaritons in transition metal dichalcogenides \cite{shan2023second}, topologically protected condensates \cite{harder2021coherent}, organic polariton condensates \cite{putintsev2024photon} as well as photon Bose-Einstein condensate \cite{schmitt2014observation}. This behavior is independent of the applied magnetic field, and most importantly consolidates the emergence of a macroscopically coherent light-matter wave in our system in both magnetic orders.

To theoretically model the behavior of the second-order coherence function, we apply the Lindblad master equation formalism to describe the polariton system and numerically investigate the coherence function with quantum Monte Carlo method (see Supplementary Note S5 for the details). In Supplementary Figs. S7c-S7f, we show the 'numerical measurement' averaged result of $g^2(\tau)$ as a function of delay time for AFM and FM scenarios with two different pumping intensity where the error bars label the corresponding standard derivation. In each figure, we observe a gradually decay of $g^2(\tau)$ from $g^2(0)$ to $1$ which reflect the thermal nature of the system.

Further, by averaging the result of $g^2(\tau)$ for a time period of 200 fs and 5 ps to account for the lower time limit as the pulse duration in experiments and the upper limit of a trial exciton reservoir lifetime \cite{dirnberger2026exciton}, respectively, we can retrieve the simulation results of $\overline{g^2(0)}$. In Fig.~\ref{figure_5}d, we present the calculated $\overline{g^2(0)}$ as the function of pumping intensity, which are in qualitative agreement with our experimental observations summarized in Fig.~\ref{figure_5}c.

\section{Conclusion}
Our work compiles clear fingerprints of the emergence of a coherent condensate of exciton-polaritons in the magnetic van der Waals crystal CrSBr. The condensate is distinct via its power-dependent threshold, its first-order spatio-temporal coherence and second-order quantum coherence. We find a new kind of magnetically tunable non-linearity arising from the strong coupling of excitons with the magnetic order in the material, which pinpoints the importance of the exciton non-linearity and especially the coupling to incoherent magnons for the global behavior of the condensate. Our work is of large interest for experiments seeking to control magnetism with cavity photons, and especially for interfacing coherent condensates with magnetically ordered phases. It enables further studies of new physical phenomena, such as the dynamic coupling of exciton-polariton condensate with the propagating coherent magnons, and polaritonic fluid dressed with skyrmion and vortice textures. Besides the conventional magnetic transport and tunneling structures \cite{chen2024twist,boix2024multistep,telford2020layered}, cavity-mediated devices based on CrSBr can fill a gap in our ever-growing inventory of novel quantum nanophotonic applications such as magnetically controllable and highly non-linear polaritonic Josephson junctions \cite{abbarchi2013macroscopic,lagoudakis2010coherent}.

\textbf{Acknowledgements.}
 C.S., B.H. and L.L. acknowledge funding from the Deutsche Forschungsgemeinschaft (DFG) in the framework of SPP 2244 (funding number: Schn1376/14-2). 
 V.S. acknowledges DFG (funding number: INST 184/222-1). 
 B.H. acknowledges the Alexander von Humboldt-Stiftung for the fellowship grant, and supportings from National Natural Science Foundation of China (NSFC) under Grants No.~12304012 and No.~12674040, and Jilin Provincial Science and Technology Development Project under Grant No.~20240602104RC. 
 M.E. acknowledges funding from the University of Oldenburg through a Carl von Ossietzky Young Researchers' Fellowship. 
 F.E. acknowledges support by DFG SFB 1375 (NOA) and BMBF FKZs 1CKISQ087K and 13XP5053A. 
 K.W.S. and O.K acknowledge the support from UK EPSRC grant EP/X017222/1. 
 Z.S. and J.R. were supported by ERC-CZ program (project LL2101) from Ministry of Education Youth and Sports (MEYS) and by the project Advanced Functional Nanorobots (reg. No. CZ.02.1.01/0.0/0.0/15\_003/0000444 financed by the ERDF).
 M.S. acknowledges the supporting from the R\&D Program of Beijing Municipal Education Commission (KM202410005011).
 I.G.S. is supported by the National Natural Science Foundation of China (NSFC) under Grant No.~W2532001.
 A.B. and I.G.S. were supported by the 2025 GTIIT-Technion Seed Grant.

\textbf{Author contributions.} C.S. conceived of the experiments. Z.S and J.R. synthesized the CrSBr crystals. F.E. and M.E. prepared and manufactured the mesa and DBR mirrors. B.H. and H.S. assembled the sample and cavity. V.S. and B.H. measured the atomic force microscopy. B.H., H.S. and L.L. performed the measurements. C.S. supervised the experiments. B.H. analyzed the experimental data with assistance from H.S.. K.W.S. and O.K. provided theoretical interpretations. M.S., A.B., and I.G.S. provided the theoretical modeling of the temporal coherence function. All authors discussed the results. B.H., C.S., M.S. K.W.S. and O.K. wrote the manuscript with inputs from all authors.

\textbf{Competing interests.} The authors declare no competing interests.

\textbf{Data Availability.} The data can be acquired upon request by contacting the Christian Schneider or Bo Han.

\section{Methods\label{methods}}

\textbf{CrSBr crystal synthesis.}
CrSBr crystals were synthesized through the direct reaction from the elements using chemical vapor transport method. High-purity chromium (99.99 \%, -60 mesh, Chemsavers, USA), bromine (99.9999 \%, Sigma-Aldrich, Czech Republic), and sulfur (granules, 99.9999 \%, Stanford Materials, USA) were combined in stoichiometric ratio within a quartz ampoule (35$\times$220 mm) corresponding to 15 g of CrSBr. An excess of 0.5 g bromine was employed to enhance vapor transport. The material was pre-reacted within an ampoule utilizing a crucible furnace gradually heated on 400°C, 500°C, 600°C and 700 °C for 24 hours at each step, while the second end of the ampoule was kept below 250 °C. Subsequently, the ampoule was positioned within a horizontal two-zone furnace to facilitate crystal growth. Initially, the temperature of growth zone was heated to 900 °C, while the source zone was heated to 700 °C for 25 hours. For the growth, the thermal gradient was reversed and the source zone was heated on 900 °C and the growth zone on 800°C over a period of 10 days. The crystals with dimensions of up to 3$\times$10 mm² were removed from the ampoule in an Ar glovebox.

\textbf{Transfer matrix simulation.}
We consider the normal incidence with a unity field intensity. The transfer matrix $T(l)$ across a dielectric layer of thickness \textit{l} is 
\begin{equation}
     T(l)=\left[\begin{matrix}
         \cos{kl} & \frac{i}{\tilde{n}}\sin{kl}\\ 
         i\tilde{n}\sin{kl} & \cos{kl}\\ 
    \end{matrix} \right],
    \label{TMM_layer}
\end{equation}
where $k$=$2\pi\tilde{n}/\lambda$ is the wavevector and $\tilde{n}(\lambda)$=$n$+$i\kappa$ is the wavelength dependent complex refractive index of the homogeneous layer material. The real and imaginary parts of the refractive index of CrSBr are taken from a previous work \cite{dirnberger2023magneto} and presented in Fig. S1e. The total transfer matrix over a structure containing \textit{m} layers is $T_{\mathrm{tot}}$=$\prod_{i=1}^{m} T_i$
, where $T_i$ is the transfer matrix of the \textit{i}th layer in the structure. The reflectivity and transmission of the whole structure are
\begin{equation}
     R=\bigg|\dfrac{\tilde{n}_{sub}(T_{11}+T_{12})-(T_{21}+T_{22})}{\tilde{n}_{sub}(-T_{11}+T_{12})+(T_{21}-T_{22})}\bigg|^2,
    \label{reflectivity}
\end{equation}
\begin{equation}
     T=\tilde{n}_{sub}\bigg|\dfrac{2(T_{12}T_{21}-T_{11}T_{22})}{\tilde{n}_{sub}(-T_{11}+T_{12})+(T_{21}-T_{22})}\bigg|^2,
    \label{transmission}
\end{equation}
where $\tilde{n}_{sub}$ is the refractive index of the semi-infinite substrate material (SiO$_2$) underneath the bottom DBR, and $T_{jk}$ is the element on the \textit{j}th row and \textit{k}th column of the transfer matrix $T_{\mathrm{tot}}$. The absorption in Fig. S1 can thus be obtained via \textit{A=1-R-T}.

\textbf{Coupled oscillator model.}
In this extended coupled oscillator model\cite{savona1995quantum,kkedziora2024non}, the coupling matrix in \ref{coupled_oscillator} includes six self-hybridized polaritons and the ground transverse cavity modes from three longitudinal mode orders. The additional two transverse modes ($E_{c{_1}}$ and $E_{c{_3}}$) are taken into account for the unconventional crossing through the self-hybridized P$_6$ polaritons at very small or very large detunings, as shown in Figs. 1d and 1e. The off-diagonal terms represent the coupling strengths between the transverse modes and self-hybridized polaritons. The energy of the transverse cavity modes is a function of the applied DC Voltages. Diagonalizing this matrix allows us to extract the dispersions of the new exciton-polariton modes as a function of the DC Voltage for the CrSBr flake in our tunable microcavity. Other higher energy polariton modes can also be fitted separately with this coupling matrix by considering transverse modes with same symmetry. We note that this extended coupled oscillator model not only yields fully consistent results as the transfer matrix method in Fig. S1g, but also provides us with the information on the Hopfield coefficients that are paramount for analyzing the polaritonic non-linearity. 

\begin{equation}
     \left[\begin{matrix}
         E_{c{_1}}\!&0\!&0\!&\frac{V_1}{2}\!&\frac{V_2}{2}\!&\frac{V_3}{2}\!&\frac{V_4}{2}\!&\frac{V_5}{2}\!&\frac{V_6}{2}\\ 
        0\!&E_{c{_2}}\!&0\!&\frac{V_1}{2}\!&\frac{V_2}{2}\!&\frac{V_3}{2}\!&\frac{V_4}{2}\!&\frac{V_5}{2}\!&\frac{V_6}{2}\\
        0\!&0\!&E_{c{_3}}\!&\frac{V_1}{2}\!&\frac{V_2}{2}\!&\frac{V_3}{2}\!& \frac{V_4}{2}\!&\frac{V_5}{2}\!&\frac{V_6}{2}\\
        \frac{V_1}{2}\!&\frac{V_1}{2}\!&\frac{V_1}{2}\!&E_{P_1}\!&0\!&0\!&0\!&0\!&0\\ 
        \frac{V_2}{2}\!&\frac{V_2}{2}\!&\frac{V_2}{2}\!&0\!&E_{P_2}\!&0\!&0\!&0\!&0\\
        \frac{V_3}{2}\!&\frac{V_3}{2}\!&\frac{V_3}{2}\!&0\!&0\!&E_{P_3}\!&0\!&0\!&0\\
        \frac{V_4}{2}\!&\frac{V_4}{2}\!&\frac{V_4}{2}\!&0\!&0\!&0\!&E_{P_4}\!&0\!&0\\
        \frac{V_5}{2}\!&\frac{V_5}{2}\!&\frac{V_5}{2}\!&0\!&0\!&0\!&0\!&E_{P_5}\!&0\\
        \frac{V_6}{2}\!&\frac{V_6}{2}\!&\frac{V_6}{2}\!&0\!&0\!&0\!&0\!&0\!& E_{P_6}\\
    \end{matrix} \right]
    \label{coupled_oscillator}
\end{equation}

\textbf{Cryogenic open cavity preparation.} 
The indentation into the mesa surface are 300 nm deep for all of the concave structures whose diameters vary from 3 $\mu$m to 6 $\mu$m. The 6 $\mu$m lens was used throughout the entire study. These indentations were etched by focused Gallium ion beam lithography (FEI Helios 600i). DBR mirrors were then sputtered on top of the mesa and silica wafer, respectively. The DBR on the mesa contains 8 pairs of 157 nm SiO$_2$ and 99 nm TiO$_2$, yielding a stop band center at 1.3745 eV (902 nm). The bottom DBR mirror contains 8 pairs of 181 nm SiO$_2$ and 111 nm TiO$_2$, yielding a stop band center at 1.2155 eV (1020 nm). The CrSBr flake was exfoliated and transferred on the bottom DBR mirror by a dry-stamping method \cite{castellanos2014deterministic}. The thickness of the flake was measured in an atomic force microscope (WITec: alpha300 RA) in tapping mode. The entire open optical microcavity was submerged in a heat-exchange helium gas that has a pressure of 20 mbar at room temperature. The microscope images of mesa and CrSBr flake are compiled in  Fig. S1. The open cavity was then loaded into a closed-cycle cryostat (attocube: attoDRY1000). All of the experiments were performed at a sample temperature of 3.5 K. The piezo-based nano-positioners have a sub-nanometer detuning increment and are ultra-stable to maintain a certain cavity length.

Fig. S1e contains the reflection of an empty cavity with a gap of 4025 nm, simulated by the transfer matrix method. For this geometry, two sharp cavity modes present in the energy range of the self-hybridized polaritons. The mode at 1.353 eV has a Q-factor of 5700, while the other mode at 1.249 eV has a Q-factor of 8600. Although the simulated quality factor (Q-factor) by the transfer matrix methods for our symmetrical dielectric cavity structure is above 5000, the measured Q-factor$\sim$1100 of the transverse modes is substantially smaller than the simulated result. The discrepancy is caused by the vibration of the cavity length at a same frequency $\sim$1.4 Hz as the closed-cycle helium pulses. As a result, the cavity resonance is broadened up to $\sim$1.1 meV, as shown in Figs. 3d and 3g. We note that the reduced Q-factor is still much larger than those of the cavity structures in previous works (Q-factor$\sim$300) using the hybrid of metallic and dielectric mirrors \cite{dirnberger2023magneto,wang2023magnetically}. It also does not impede us from the strong light-matter coupling regime and the exciton-polariton condensate.

\textbf{Optical microscopy.} The optical setup is in a confocal geometry. For the PL measurements in Figs. \ref{figure_1}c, Figs. \ref{figure_3}a-\ref{figure_3}c and magneto-PL in Fig. S2, we used a 725 nm laser excitation with a 5 ps pulse width and 78 MHz repetition rate, which was generated from a supercontinuum white light laser and an acousto-optic tunable filter (NKT Photonics: SuperK SELECT). For the pump-power dependent measurements, the 725 nm laser excitation of 200 fs pulse width and 76 MHz repetition rate is generated from a mode-locked Ti-sapphire laser (Coherent: Mira 900 Femtosecond).  Laser excitation and signal collection were realized by a long work-distance lens objective (Thorlabs: 354105-B, NA=0.6, f=5.5 mm) that focused on the surface of the top DBR. The static PL signals were transmitted in free-space, collected and focused by a plano-convex lens (f=450 mm) on the spectrometer slit (Andor: Shamrock SR-500i). The signals were then dispersed by a 600 mm$^{-1}$ grating and recorded by a charge coupled device (CCD, Andor: iKon-M 934). 

Fig. S1f shows that the PL emission of a LPB mode of CrSBr in the external cavity is linearly polarized along the crystal b-axis. The polarizers on the excitation and detection can be aligned to the crystallographic b-axis by searching for the maximum PL emission intensity. A half waveplate on the detection side rotates the polarization of the signal to realize the polarization-resolved measurements. The fit yields a unity polarization degree along the crystallographic b-axis. 

To extract the polariton effective temperature, we analyze the emission intensity in Fig. \ref{figure_1}e by leveraging the intrinsic properties of the material. Since CrSBr excitons exhibit a highly anisotropic or a quasi-1D dispersion \cite{liebich2025controlling}, the resulting polaritonic modes within the rotationally invariant 2D planar cavity inherit this distinct anisotropy. Consequently, the emission intensity polarized along the crystal b-axis (Fig. \ref{figure_1}e) directly reflects the polariton occupancy. As for the polaritonic non-linearity measurements using 0D discrete cavity modes, to exclude the potential anisotropic excitation effects, we align the pump laser polarization to the middle between the crystallographic a and b axes of the CrSBr flake. As expected, due to the reduction of the power projected along the crystal b-axis, the P$_{th}$ value in 0D cavity is higher than that in the 2D planar cavity.

\textbf{First order correlation measurements.} 
First-order correlation was measured via a Michelson interferometer. We firstly used the spectrometer to disperse the static PL signal and filtered the spectral window of the LPB by using edge filters (Semrock: TLP01-995 and TSP01-995). The interferometer then divided the filtered signal into the reference and delay arms. The delay arm has a silver mirror on a kinematic mount, while the reference arm has a retroreflector (Thorlabs: PS976M) that reflects back a spatially displaced and inverted image. The reflections of two arms were collected by the same plano-convex lens (f=450 mm) to focus the signals on the spectrometer slit. The grating was then set to zero-order to check the real-space emissions from the reference and delay arms. We adjusted the angle of the delay arm mirror to spatially overlap the signals on the CCD. The zero-delay position was affirmed by the maximum interference fringe visibility. 

We then recorded the excitation power dependence at zero delay of the delay arm, reference arm and interference images. $g^{(1)}(\vec{r},\Delta\tau)2\cos(\Delta\phi)$ can be readily obtained from the captured interference, delay and reference arms images via Eq. (\ref{1st_order_correlation}). To remove the phase factor, we applied the two-dimensional fast Fourier transform (FFT) on the real space $g^{(1)}(\vec{r},\Delta\tau)2\cos(\Delta\phi)$ (using WaveMetrics Igor Pro), yielding two Fourier peaks in the momentum-space. One Fourier peak was then filtered and displaced to the center ($k_x$=0, $k_y$=0) of the momentum-space. A following inverse fast Fourier transform (IFFT) brought the complex $g^{(1)}(\vec{r},\Delta\tau)$ back to the real-space. The amplitude of the spatially dependent $g^{(1)}(\vec{r},0)$ was then calculated for the condensate in AFM and FM orders, and showed in Figs. 4b and 4c, respectively.

\textbf{Second order Correlation measurements.}  
The second order correlation was measured in a Hanbury Brown and Twist interferometry. The filtered signal was then collected by a fiber-coupled zoom collimator (Thorlabs: ZC618APC-B), whose output coupled with one input channel of a 2$\times$2 fiber beam splitter (Thorlabs: TW850R5A2), while the second input was idle. The outputs of the fiber beam splitter connected to two same APDs (Laser components: Count T) whose electronic output was sent to the time correlator (Swabian instruments: Time-Tagger 20). A 250 ps binning width was used for these time-correlated measurements to match the temporal resolution (350 ps). The experimental $\overline{g^{2}(0)} = N_0/N_S$ values presented in Fig. 5c were extracted 
by normalizing the integrated intensity ($N_0$) of the central peak at $\tau=0$ to the  averaged integrated peak intensities ($N_S$) of ten side peaks at $\tau\neq0$.

\setupSI

\begin{figure*}[h]
\includegraphics[width=0.85\columnwidth]{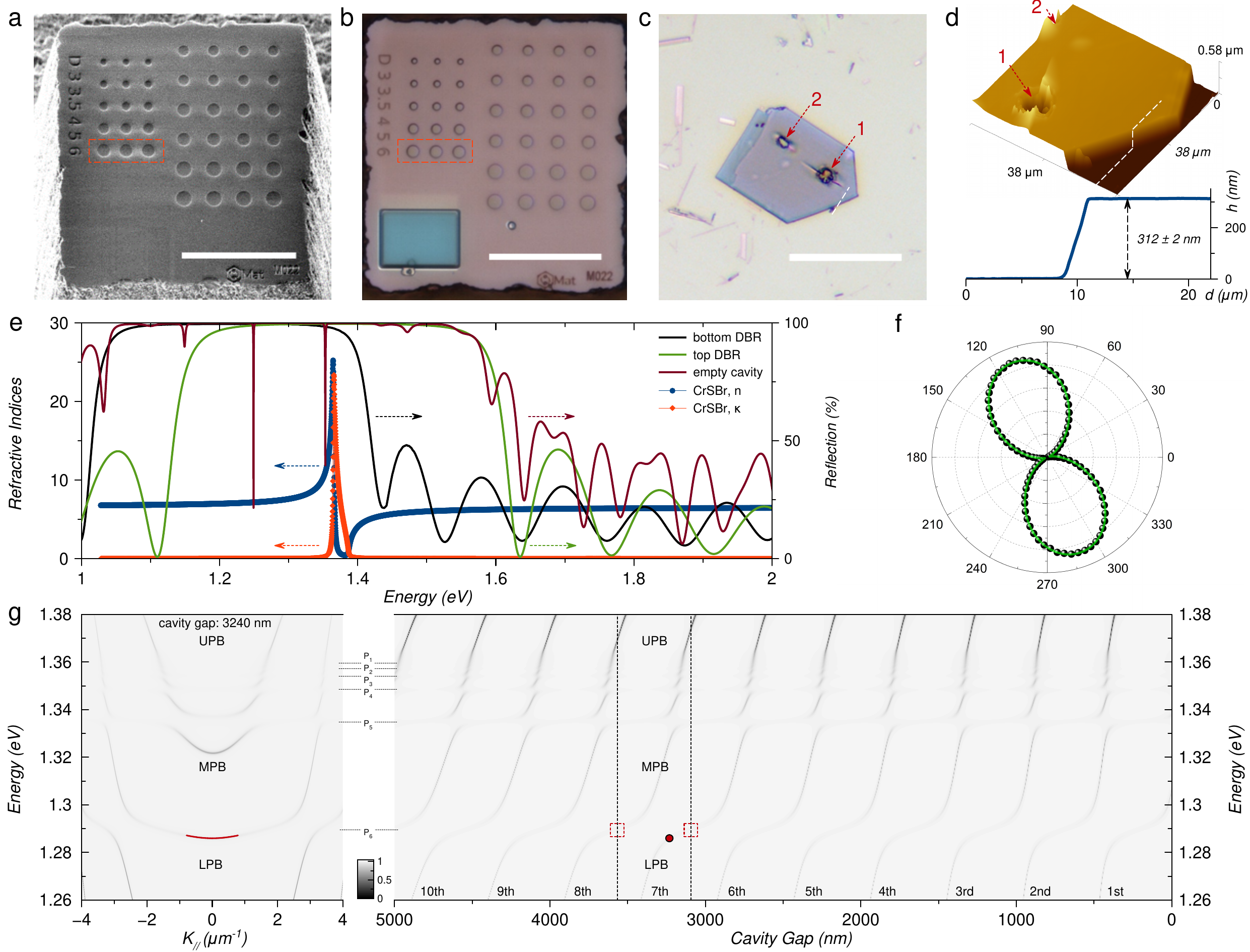}
   \caption{\textbf{Cavity properties.} \textbf{a.} Scanning electron microscope image of the mesa after FIB etching. \textbf{b.} Optical microscope of the mesa with sputtered DBR. The 6 $\mu$m lenses (concave DBR) are marked by the red frame. \textbf{c.} Microscope image of the CrSBr flake transferred on the bottom DBR. The scale bars in \textbf{a-c} are all 50 $\mu$m. The holes (1: 6 $\mu$m$\times$ 4 $\mu$m; 2: 4 $\mu$m $\times$ 5 $\mu$m) are burned through the 6 $\mu$m lenses by 725 nm fs laser at a power of 0.92 nJ/pulse. \textbf{d.} Upper panel: atomic force microscopy of the CrSBr flake in \textbf{c}. Lower panel: step profile along the white dashed lines in the upper panel. The thickness (h=312$\pm$2 nm) matches perfectly with the simulation by the transfer matrix in Fig. 1c. \textbf{e.} Reflection of the top and bottom DBRs, and empty cavity with a gap of 4025 nm, which are simulated by the transfer matrix method, and the complex refractive index \textit{$\tilde{n}$$=n+i\kappa$} of CrSBr used for the transfer matrix simulation in Fig. 1c . \textbf{f.} Polarization dependent PL intensity (black) of the LPB at DC=20 V in Fig. \ref{figure_S2}a . The fit (green) determines crystallographic b-axis as 110° (290°) in our experimental geometry. \textbf{g.} Transfer matrix simulated reflection of a 312 nm CrSBr in a planar cavity with same configurations as experiments. Right panel: only normal incidence is considered, where in-plane photon wavevector $K_{//}=0$. The black dashed lines mark the strong coupling for the 7th longitudinal mode in the external cavity, corresponding to the experiments in Fig. 3d. The red boxes around 3650 nm and 3000 nm cavity gaps mark comparable DC voltages of 0 and 60 V, respectively. For each mode order, the LPB connects to the MPB with a lower mode order, fully consistent with the PL and 9$\times$9 coupled oscillator model in Fig. 3d. The self-hybridized polaritons are marked. Left panel: polaritonic dispersions as a function of $K_{//}$ with a 3240 nm cavity gap marked by the red dot in the right panel, which is same to DC=44 V in Fig. 3d. Fitting the LPB dispersion (red curve) yields an effective mass $m_{eff}^{LPB}$=2$\times$10$^{-5}m_{e}$, where $m_e$ is the free electron mass.}
   \label{figure_S1}
\end{figure*}

\section{Properties of sample position 1 and 2}
\label{Supplementary note 2}

\begin{figure*}[t]
\includegraphics[width=\columnwidth]{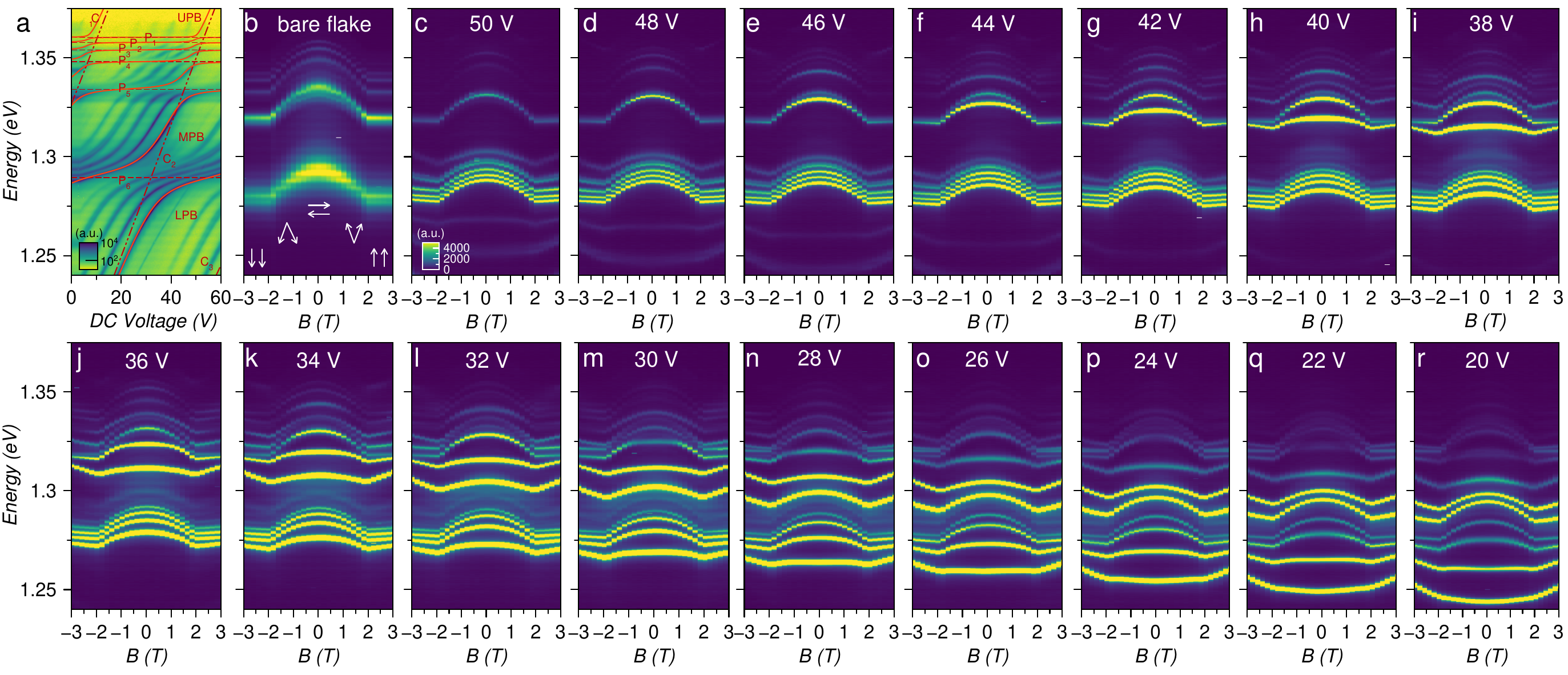}
    \caption{\textbf{Magneto-PL of sample position 2 at different cavity detunings.} \textbf{a.} Cavity detuning PL in AFM order (0 T). The results are similar to sample position 1 in the main text with slightly different energies of the self-hybridized polaritons (see Table.  \ref{table_1}). \textbf{b-r.} Magneto-PL of the bare flake (same as Fig. 1c) and new polariton modes that are performed between ± 3 T for detuning DC voltages from 20 V to 50 V in \textbf{a}. All graphs share the same colorbar in \textbf{c}. Note that the cavity length of our setup gets smaller and the modes slightly blueshift in magnetic fields. These drifts are more obvious for modes with higher photonic components, which are less than 1 meV below the out-of-plane saturation magnetic field $|B|\sim2$ T.}
    \label{figure_S2}
\end{figure*}

As shown in Fig. \ref{figure_S1}c, both sample positions are burned accidentally in power dependent measurements. Sample position 2 also exhibits six self-hybridized polariton states as position 1. Fig. \ref{figure_S2} presents the magneto-PL measurements on sample position 2 with different cavity detunings. If not mentioned, all the power-dependent measurements in the main text are performed on sample position 1. The energies of the self-hybridized polaritons ($E_{P{_1}}$-$E_{P{_6}}$) and the coupling strengths ($V_1$-$V_6$) used in the fitting are summarized in Table. \ref{table_1}.The coupling strengths of the self-hybridized polaritons on sample position 1 are nearly constant for both phases.

\begin{widetext}
\begin{center}
\begin{table}[ht]
\begin{tabular}{|c| c c c c c c | c c c c c c|} 
 \hline
   Phase\textbackslash unit: eV & $E_{P{_1}}$ & $E_{P{_2}}$ & $E_{P{_3}}$ & $E_{P{_4}}$ & $E_{P{_5}}$ & $E_{P{_6}}$ & $V_1$ & $V_2$ & $V_3$ & $V_4$ & $V_5$ & $V_6$\\ [0.5ex] 
 \hline\hline
 Position 1 (forced FM)  & 1.3515 & 1.3482 & 1.3442 & 1.3376 & 1.3248 & 1.2827 & 0.0040 & 0.0046 & 0.0052 & 0.0090 & 0.0164 & 0.0360 \\ [0.5ex] 
 \hline
  Position 1 (AFM) & 1.3660 & 1.3634 & 1.3597 & 1.3540 & 1.3395 & 1.2944 & 0.0040 & 0.0050 & 0.0056 & 0.0090 & 0.0164 & 0.0374 \\
 \hline
 Position 2 (AFM) & 1.3604 & 1.3578 & 1.3540 & 1.3480 & 1.3341 & 1.2895 & 0.0030 & 0.0046 & 0.0056 & 0.0080 & 0.0154 & 0.0330 \\ 
 \hline
\end{tabular}
\caption{Self-hybridized polariton energies ($E_{P{_1}}$-$E_{P{_6}}$) and their coupling strengths ($V_1$-$V_6$) in the external cavity.}
\label{table_1}
\end{table}
\end{center}
\end{widetext}

\section{Polariton non-linearity measurements for different cavity detunings and magnetic orders}

\begin{figure*}[t]
\includegraphics[width=2\columnwidth]{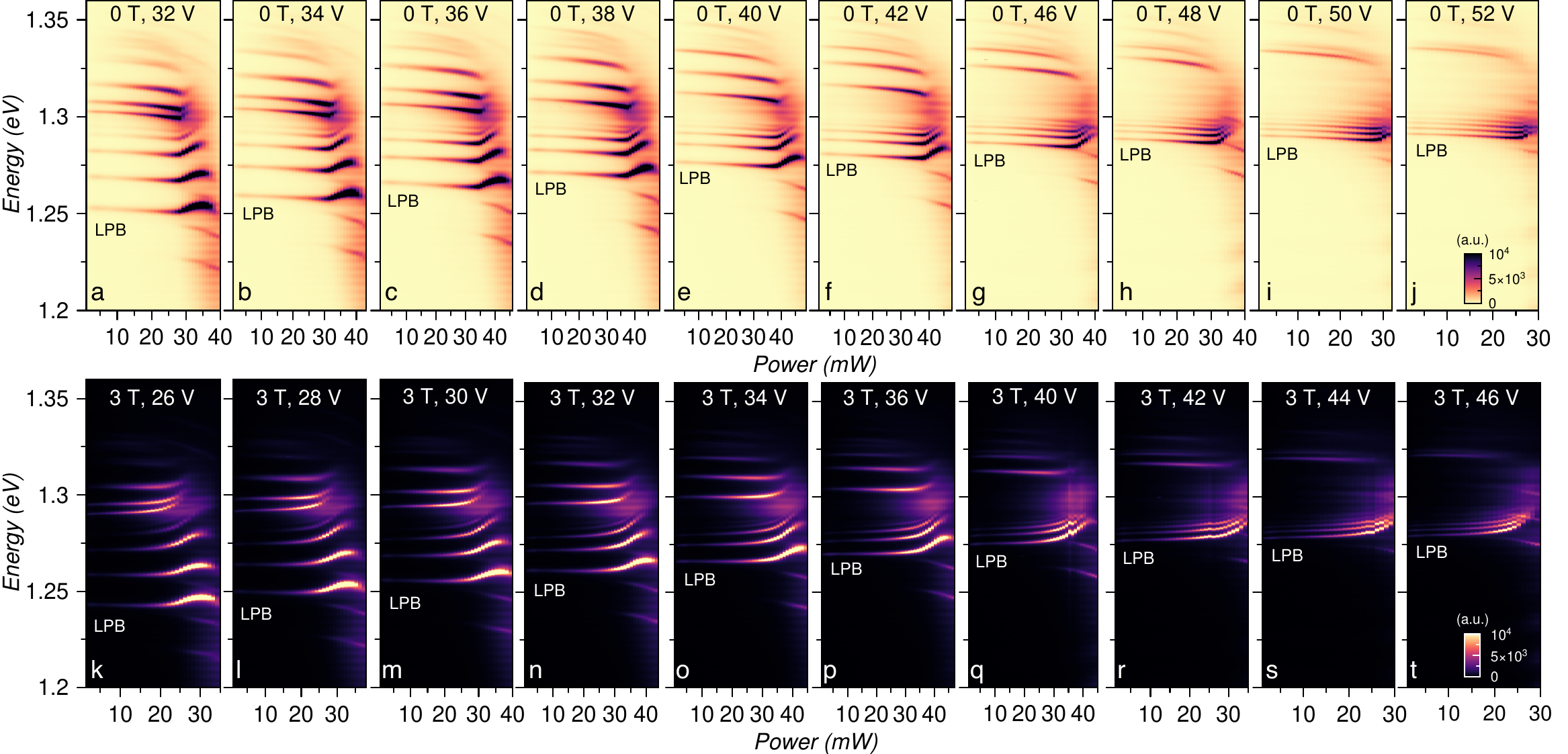}
   \caption{\textbf{Power dependent PL of sample position 1.} \textbf{a-j.} cavity detunings for DC=32-52 V and AFM order (0 T). \textbf{k-t.} cavity detunings for DC=26-46 V and FM order (3 T). The two colorbars apply to measurements with contrasting magnetic orders. The DC voltages in \textbf{a-j} and \textbf{k-t} correspond to those marked in cavity detuning PL measurements in Figs. 3a and 3b, respectively. The power dependent PL at DC=44 V (0 T) and DC=38 V (3 T) are shown in Fig. 3d and Fig. 3e, respectively.}
    \label{figure_S3}
\end{figure*}

Polaritonic non-linearities with different cavity detunings and magnetic orders are obtained via power dependent PL measurements on sample position 1 using the 0D cavity structure. For each study, we keep a certain cavity detuning at a constant DC voltage between 32-52 V for the AFM order at 0 T, and 26-46 V for the FM order at 3 T. The experimental results are compiled in Fig. \ref{figure_S3} and two exemplary cases in Figs. 3d and 3e in the main text. The energy shifts of the LPBs in experiments are then fitted and summarized in Figs. S4a and S4b. 

\begin{figure*}[t]
\includegraphics[width=2\columnwidth]{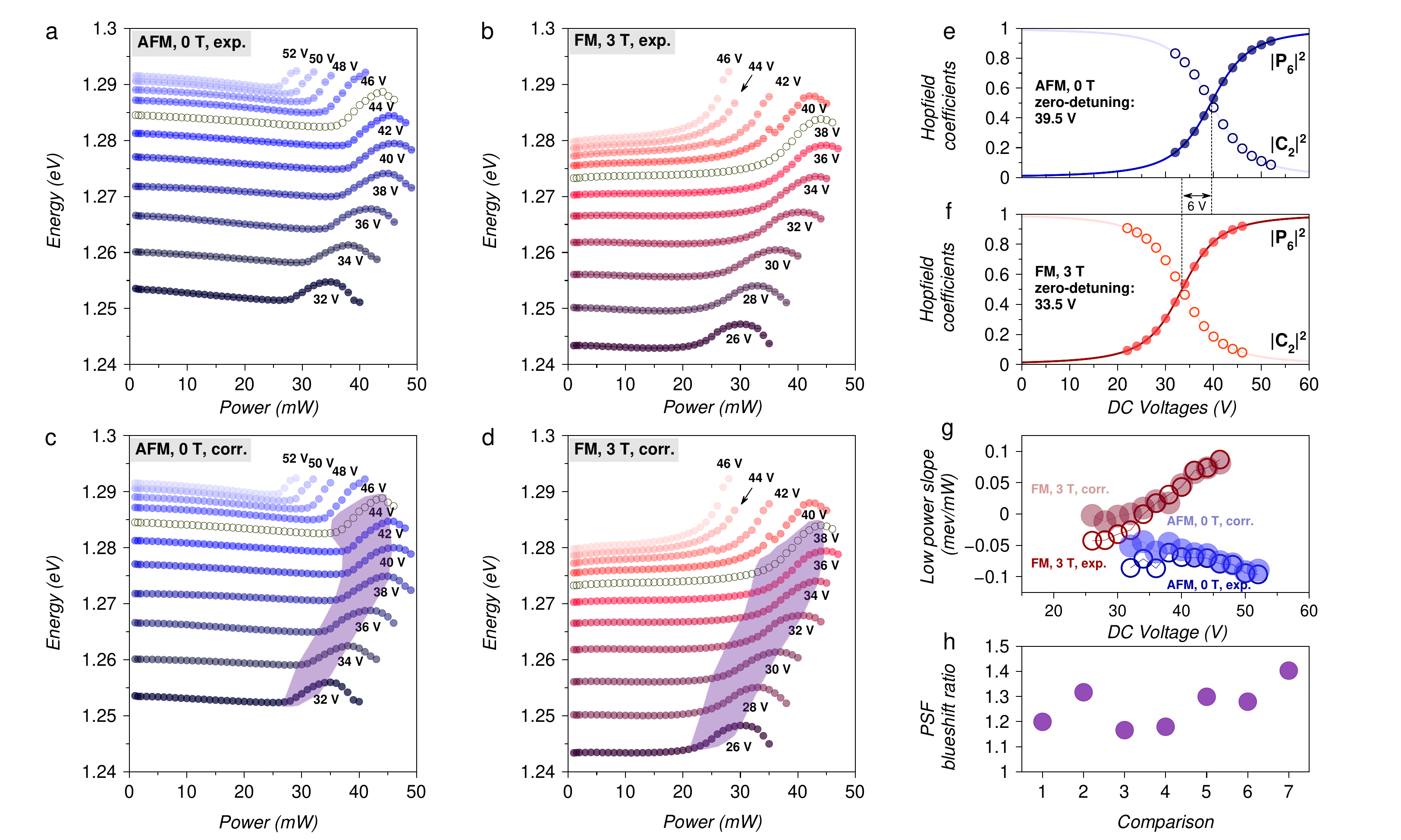}
   \caption{\textbf{Polaritonic non-linearity of sample position 1 with various cavity detunings and two magnetic orders.} Experimental LPB energy shifts with increasing pump power in \textbf{a.} AFM order (0 T) and \textbf{b.} FM order (3 T), fitted from the raw data in Fig. \ref{figure_S3}. Corrected LPB energy shifts with increasing pump power in \textbf{c.} AFM order and \textbf{d.} FM order. \textbf{e.} AFM order and \textbf{f.} FM order Hopfield coefficients of $P_6$ self-hybridized polariton $|P_6|^2$ and the external cavity mode $|C_2|^2$ of the LPBs with different external cavity detunings, calculated by a 2$\times$2 coupled oscillator model in Eqs. S1-S4. The dots and circles mark the cavity detunings in Figs. 3d and 3e of the main text where the polariton non-linearities are measured. A 6 V difference in zero-detunings of FM and AFM orders are indicated. \textbf{g.} Experimental (empty symbols) and corrected (filled symbols) slopes of the LPB energy shifts below 20 mW in \textbf{a-d}, fitted by a linear function. \textbf{h.} Ratio of non-linear phase space filling (PSF) blueshift between FM and AFM spin orders. The data in the shaded regions of 32-44 V in \textbf{c} and 26-38 V in \textbf{d} are considered. Each comparison takes a 6 V difference in DC voltages to account for a same detuning condition of LPB in AFM and FM magnetic orders. \textbf{i, j} Power dependent LPB intensities of LPB for AFM order (DC=44 V) and FM order (DC=38 V) in double logarithm scale (see linear scale plots in Figs. 3f and 3i of the main text).
   }
    \label{figure_S4}
\end{figure*}

For all detunings in the AFM order, the LPB firstly experiences a moderate linear redshift, and then a giant blueshift follows until the condensate depletes. However, for the detunings in the FM order, the energy slope of LPB at low pump powers changes sign from moderate redshift to moderate blueshift, when the detuning voltage is tuned above 36 V. And at high pump powers, the giant blueshift in FM order is similar to that in AFM order. By using a linear fit of the LPB energy shifts below 20 mW, the energy slopes (empty spherical symbols) for different detuning scenarios and magnetic orders are obtained and summarized in Fig. \ref{figure_S4}g. 

We can see that the energy shifts in low power range of the LPB in AFM and FM orders present obviously opposite trends. As the LPB becomes more excitonic (at higher DC voltages), the redshift slope of the LPB in the AFM order becomes smaller (larger amplitude), while in the FM order the slope keeps increasing. This magnetic order dependent phenomena have been observed for the magnetic excitons in CrSBr as a result of coupling to the incoherent magnons that are excited by the temperature \cite{dirnberger2023magneto}. In addition, we note that the redshift should not exist for the highly photonic LPB (DC=26 V) in the FM order because of the much reduced excitonic components and correspondingly little interaction between excitons and incoherent magnons. This effect is actually caused by the redshift of the cavity modes due to the thermal expansion of our open cavity as we ramp up the pump power and increase the condensate density. In Supplementary Fig. \ref{figure_S5}b, we can see the 725 nm pump energy are much more localized in the top DBR, which is the reason of cavity energy drift by heating up the top mirror. Another proof of cavity thermal expansion is that the extrapolation of the slope value in the AFM order towards the pure photonic regime converges with the FM scenario at around -0.05 meV/mW, which purely comes from the cavity thermal drift.

To quantify the redshift due to the cavity drift, we need to know the photonic Hopfield coefficients that are specifically contributed by the external cavity modes. Figures. \ref{figure_S4}e and \ref{figure_S4}f show the Hopfield coefficients of the LPB mode in AFM and FM orders, respectively, which are calculated by using a 2$\times$2 coupled oscillator model by considering only the P$_6$ self-hybridized polariton mode and the C$_2$ cavity mode:
\begin{equation}
     \left(\begin{matrix}
        E_{c_2}\!&V_6/2\\ 
        V_6/2\!&E_{P_6}\\
    \end{matrix} \right)
    \label{coupled_oscillator_2}
\end{equation}
\begin{equation}
E_{\mathrm{LPB}}=\dfrac{1}{2}[E_{c_2}+E_{P_6}-\sqrt{V_6^2+{(E_{c_2}-E_{P_6})}^2}]
\end{equation}
\begin{equation}
|P_6|^2=\dfrac{1}{2}+\dfrac{E_{c_2}-E_{P_6}}{2\sqrt{(E_{c_2}-E_{P_6})^2+V^2}}
\label{X_hopfield}
\end{equation}
\begin{equation}
|C_2|^2=\dfrac{1}{2}-\dfrac{E_{c_2}-E_{P_6}}{2\sqrt{(E_{c_2}-E_{P_6})^2+V^2}}
\end{equation}
This model is valid for two reasons. Firstly, the energy difference between P$_5$ and P$_6$ intrinsic self-hybridized polaritonic modes is larger than the coupling strength $V_6$, so that the strong coupling of $P_6$ to the external cavity modes can be regarded as independent of P$_1$-P$_5$ self-hybridized polariton resonances. Secondly, the usage of two additional cavity modes in our 9$\times$9 coupled oscillator model in Figs. 3a and 3b only becomes vital for unconventional polariton dispersion in either highly photonic (DC< 5 V) or highly excitonic (DC>55 V) cases, so that utilizing only the middle cavity mode '$c_2$' has negligible influence to the cavity detunings where the polaritonic non-linearities are measured (Fig. \ref{figure_S3})

\begin{figure*}[t]
\includegraphics[width=2\columnwidth]{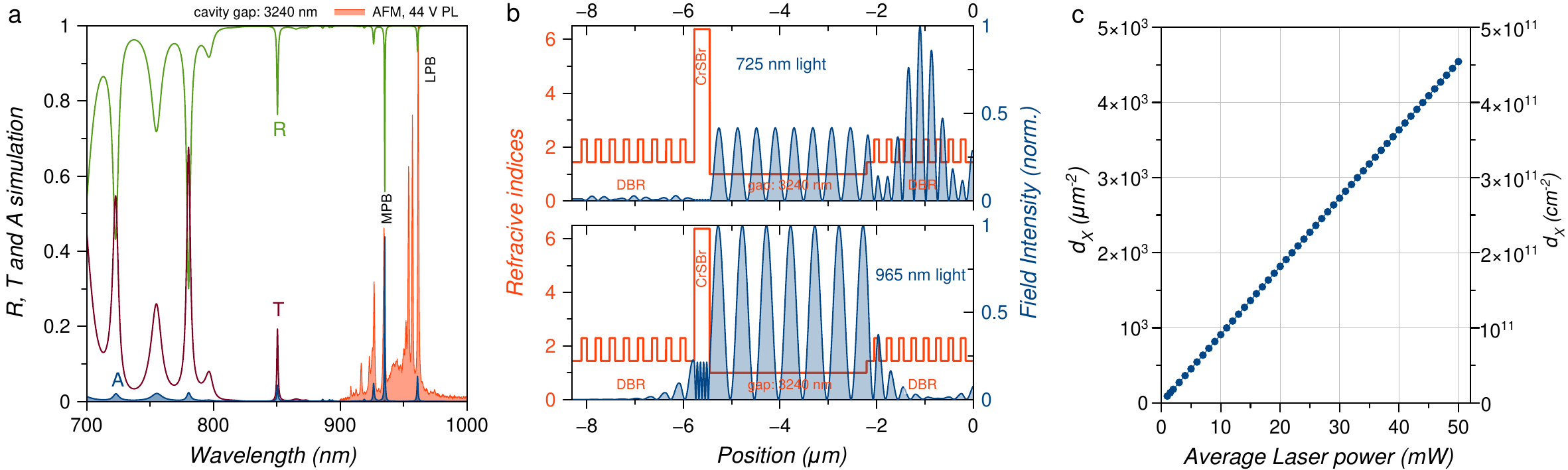}
  \caption{\textbf{Exciton density estimation.} \textbf{a.} Transfer Matrix simulation of Reflection (R), transmission (T), and absorption (A), using a cavity gap of 3240 nm as in Fig. \ref{figure_S1}g. The simulation matches very well the PL spectrum (AFM order, DC=44 V as in Fig. 3a). Some tiny discrepancy might be due to sample inhomogeneity. The absorption at 725 nm is 0.0209. \textbf{b.} Electric field intensity for 725 nm (top panel) and 965 nm (bottom panel) light in the cavity manifested by the refractive index structure. \textbf{c.} Conversion between exciton density in each layer and average pump power. Left (right) axis: unit in $\mu m^{-2}$ ($cm^{-2}$).}
    \label{figure_S5}
\end{figure*}

We note that the Hopfield coefficients in Figs. S4e and S4f only represent the proportionalities of P$_6$ self-hybridized polariton and the external 'C$_2$' cavity mode in the LPB. The excitonic and total photonic Hopfield coefficients should be renormalized by taking into account the intrinsic self-hybridization in our CrSBr flake. For example, using the methods provided in a previous work \cite{dirnberger2023magneto}, the excitonic Hopfield coefficient of P$_6$ mode with a maximum magnetic redshift of -13.3 meV is determined as $|X|_{P_6}^2$= 0.76. In presence of an external cavity, the excitonic admixture will be further diluted. For the lowest LPBs in Fig. 3c of the main text, the Hopfield coefficient of P$_6$ self-hybridized polariton is $|P_6|^2$= 0.74, which means the LPB has an excitonic Hopfield coefficient $|X|^2$=$|X|_{P_6}^2\times|P_6|^2$= 0.56 and total photonic Hopfield coefficient $|C|^2$= 0.44. These results are consistent with the $|X|^2$= 0.49 and $|C|^2$= 0.51 of the lowest LPB in a similar cavity detuning scenario (36 V) on sample position 2 in Fig. \ref{figure_S2}j, which is directly calculated by using its magnetic shift. 

In the following, we calibrate the polariton non-linearity by removing the contribution from cavity redshifts with increasing pump power. We use the experimentally extracted redshift slope $s=-0.0474$~meV/mW of the LPB mode at 22 V cavity detuning ($|C_2|^2$=0.907) of the FM order to calculate the pure cavity drift slope due to the thermal expansion: $s_0=\dfrac{s}{|C_2|^2}$=-0.0523 meV/mW. We then use $s_0$ to renormalize the cavity redshift slope of each detuning by multiplying their photonic Hopfield coefficients $|C_2|^2$. We obtain the redshift caused by the cavity thermal expansion $\delta=s_0P$, where $P$ is the pump power. The LPB mode energy modified by the cavity drift is
\begin{equation}
E_{\mathrm{LPB}}'=\dfrac{1}{2}[E_{c_2}+E_{P_6}+\delta-\sqrt{V_6^2+{(E_{c_2}+\delta-E_{P_6})}^2}].
\end{equation}
The polariton redshift caused by the cavity expansion is 
\begin{eqnarray}
&&\Delta=E_{\mathrm{LPB}}'-E_{\mathrm{LPB}}\\
    =&&\dfrac{1}{2}[\delta-\sqrt{V_6^2+{(E_{c_2}+\delta-E_{P_6})}^2}+\sqrt{V_6^2+{(E_{c_2}-E_{P_6})}^2}].\nonumber
    \label{offset}
\end{eqnarray}

The experimental values of the polariton non-linearity in Figs. \ref{figure_S4}a and \ref{figure_S4}b are then corrected by removing an offset of $\Delta$ in Eq. \ref{offset}, the corrected results of which are presented in Figs. \ref{figure_S4}c and \ref{figure_S4}d. The slopes below 20 mW of the corrected data are fitted and plotted as the filled spherical symbols in Fig. \ref{figure_S4}g. Now, we see that the extrapolation of the corrected slopes in the AFM order towards purely photonic regime converges with the corrected slopes in the FM scenario at 0 meV/mW. For more excitonic regimes that are DC $>$ 44 V in AFM order and DC $>$ 38 V in FM order, the compensation effect of cavity drift is nearly negligible. In low power regime, the LPB redshifts (blueshifts) in AFM (FM) order for all cavity detunings. In high power regime, the LPBs have giant blueshifts for both magnetic orders and all cavity detuning scenarios.

\section{Exciton density estimation}

The pump is $\lambda$=725 nm laser that has a 200 fs pulse duration at a pulse-repetition frequency (PRF) of 76 MHz. The energy of a single pulse with P=1 mW measured average power is
\begin{equation}
    E^{1 mW}_{\mathrm{pulse}}=\dfrac{P}{PRF}=\dfrac{10^{-3}\ W}{76\times{10^6}\ Hz}\simeq13.16\ pJ.
\end{equation}
The energy of one photon is
\begin{equation}
\begin{aligned}
     E_{\mathrm{photon}}&=\dfrac{hc}{\lambda}\simeq2.74\times10^{-7} pJ,
\end{aligned}
\end{equation}
where $h$=$6.62607015\times10^{-34}J\cdot\textit{Hz}^{-1}$ and $c$=$2.99792458\times10^8\ m\cdot\textit{Hz}$ are the Planck constant and vacuum light speed, respectively. The number of photons in a single pulse with average powers of 1 mW and 50 mW are 
\begin{equation}
   n^{1 mW}_{\mathrm{photon}}=\dfrac{E^{1 mW}_{\mathrm{pulse}}}{E_{\mathrm{photon}}}=4.80\times10^{7}, n^{50 mW}_{\mathrm{photon}}=2.40\times10^{9}. 
\end{equation}
We utilize transfer matrix methods to simulate the open cavity structure (DBR/gap/CrSBr/DBR). Fig. \ref{figure_S5}a shows that applying a cavity gap of 3240 nm yields polariton modes matching very well with the PL spectrum of the LPB in AFM order as in Fig. 3c in the main text. As shown in the top panel of Fig. \ref{figure_S5}b, the electric field intensity of 725 nm incidence light drops substantially in the CrSBr slab and the bottom DBR, signifying much higher dielectric indices of CrSBr than the materials consisting of DBR. However, the electric field intensity of the 725 nm pump laser shows much higher intensity in the top DBR than in the cavity gap, which is the main source for the thermal redshifts of cavity modes even at high photonic regime (see hollow dots in Supplementary Fig. \ref{figure_S4}g). In contrast, the electric field intensity of 965 nm light that is around the LPB condensate show stronger intensity in the CrSBr slab and cavity gap, and much less localization in both DBRs (Fig. \ref{figure_S5}b bottom panel). 

We subsequently quantify the absorption of the cavity system at 725 nm: A=1-R-T=0.0209 in Fig. \ref{figure_S5}a. The exciton number for a 50 mW average excitation power is thus calculated as 
\begin{equation}
    n_X^{50 mW}=A\cdot n^{50 mW}_{\mathrm{photon}}\simeq5.02\times10^{7} .
\end{equation}
We exclude the excitation scenarios by more than one laser pulses. The exciton reservoir depletes completely before the arrival of a following laser pulse because the exciton lifetime $\sim$15 ps \cite{meineke2024ultrafast} in a CrSBr flake with similar thickness of 400 nm is three orders shorter than the pulse interval of 13.16 ns in our experiments. The polariton lifetime is supposed to be even shorter than the purely excitonic scenario. Considering a layer thickness of 0.8 nm \cite{rizzo2022visualizing,tschudin2024imaging}, our 312 nm CrSBr flake contains $m=390$ layers. The excitation area on CrSBr $S_X$=$\pi\cdot r^2$=$9\pi$$\mu$m$^2$ is referred from the burned regions in Fig. \ref{figure_S1}c.
For an average pump power of 50 mW, the exciton density in each CrSBr layer is thus derived as
\begin{equation}
    d_X^{50 mW}=\dfrac{n_X^{50 mW}}{m\cdot S_X}\simeq 4.53\times10^3 \mu m^{-2}.
\end{equation}
We can then rescale the exciton density of each layer for other pump powers (Fig. \ref{figure_S5}c). 

We can see that for this maximum pump power applied in our experiments, the exciton density in each CrSBr layer is well below the Mott density $n_{\mathrm{Mott}}$ $\sim10^{13} \text{cm}^{-2}$ in the transition metal dichalcogenide monolayers \cite{chernikov2015population,wang2018colloquium}. It is comparable to $n_{\mathrm{Mott}}\sim10^{11}$ $\text{cm}^{-2}$ in III-V \cite{kappei2005direct,huber2005broadband,rossbach2014high} and II-VI \cite{cain1997photoluminescence,teran2006optical} semiconductor quantum wells. However, due to the highly anisotropic reduced masses and dielectric properties, the excitonic wavefunction is quasi-1D along the b-axis with substantial charge density on Cr \textit{3d}, S and Br \textit{p} orbitals \cite{wilson2021interlayer,klein2023bulk,ziebel2024crsbr,wu2022quasi,smolenski2025large}. The exciton radii along the a-axis \cite{semina2024excitons,klein2023bulk} are on the same order of the unit cell scales \cite{smolenski2025large,liu2022three}, so that the excitons in CrSBr can be regarded as mixed Frenkel and Wannier-Mott type \cite{semina2024excitons,klein2023bulk,datta2025magnon}. 
To be more quantitative, the effective exciton Bohr radius in CrSBr ($a_B\sim$ 1.2 nm) \cite{semina2024excitons} is considerably smaller than those Wannier-Mott excitons ($a_B\geq4$ nm) in conventional III-V and II-VI semiconductor systems \cite{kavokin2017microcavities}, so that in a rough estimation \cite{klingshirn2012semiconductor} the Mott density n$_{\mathrm{Mott}}\sim$ a$_B^{-2}$ in CrSBr ought to be at least one order of magnitude higher than 10$^{11}$ $\text{cm}^{-2}$. Therefore, we suppose during the whole polariton nonlinearity measurements our system should not have experienced Mott transition, where electrons and holes are in a weak Coulomb-correlated plasma instead of bound excitons \cite{mott1949basis,mott2004metal}.

\section{Theoretical model for exciton-magnon coupling}
\subsection{Exciton}

Here, we describe a theoretical model for excitons in CrSBr samples. We consider a system with bilayer configuration, as this allows to understand the overall behaviour in the presence of interlayer hybridization and spin ordering. We stress that our analysis of the bilayer configuration is a minimal example that is also representative of bulk physics for these properties. In this case a bulk van der Waal material can be seen as a stack of quasi-2D layers with weak momentum dispersion in the out-of-plane direction, similar to consideration to other materials like cuprates.

The excitonic energy can be written as
\begin{align}
E_X(\theta_{1},\theta_{2})=&\varepsilon_g(\theta_{1}-\theta_{2})+E_b+\rho_\mathrm{X} g_\mathrm{X} ,
\label{eqn:EX}
\end{align}
where $E_b$ is the exciton binding energy, and  $\varepsilon_g$ is the energy bandgap that accounts for corrections from magnetic ordering. This is described by angles $\theta_1$ and $\theta_2$ between the external magnetic field and the spin in layer $1$ (upper) and and layer $2$ (lower). The bandgap of the material can be described as
\begin{align}
\varepsilon_g(\theta_{1}-\theta_2)=\varepsilon_0+t\cos\left(\frac{\theta_1-\theta_2}{2}\right),
\end{align}
including the dependence on the relative spin direction of the spin in upper and lower layer. This spin-dependent term originates from an overlap between interlayer spin wavefunctions. The overlap amplitude $t$ is estimated to be $12$~meV from our experiment, as shown in Fig.~3. In the presence of background exciton with density $\rho_\mathrm{X}$, this will lead to the non-linear shift of the exciton due to exciton-exciton exchange interaction. 

To theoretically estimate $g_\mathrm{X}$, we first model the exciton by the following creation field operator as
\begin{align}
    X^{\dagger}_{\sigma}=\sum_{\ell_c\ell_v}\sum_{\mathbf{k}} C_{\ell_c\ell_v}^{\sigma}\psi^{\ell_c\ell_v}_{\sigma}(\mathbf{k})a^\dagger_{\ell_c\mathbf{k}\sigma}b_{\ell_v,-\mathbf{k} {\sigma}},
\end{align}
where $\psi^{\ell_c\ell_v}_\sigma(\mathbf{k})$ is the exciton wavefunction with $\sigma$ and $\mathbf{k}$ being the spin and in-plane crystal momentum. The layer index for conduction band electron is $\ell_c$ and for valence band hole $\ell_v$. As previously mentioned, we model our system by a bilayer CrSBr ($\ell_c,\ell_v=1,2$), as this is the minimal case to investigate interlayer-intralayer hybridization effects. We note that including more layers into the theory will not change the result qualitatively.\cite{Ceferino2020} 
With this, the intralayer exciton wavefunction are $\psi^{11}_\sigma(\mathbf{k})$, $\psi^{22}_\sigma(\mathbf{k})$ and the interlayer wavefunctions are $\psi^{12}_\sigma(\mathbf{k})$, $\psi^{21}_\sigma(\mathbf{k})$. The interlayer and intralayer exciton hybridization coefficient is $C^{\ell\ell'}_{\sigma}$. The creation field operator of conduction band and the annihilation field operator of the valance band is $a_{\mathbf{k}\sigma}^\dagger$ and $b_{\mathbf{k}\sigma}$.

The intralayer and interlayer excitonic states satisfy the Wannier equation
\begin{eqnarray}
    &&[\varepsilon_c(\mathbf{k})-\varepsilon_v(\mathbf{k})]\psi^{\ell_c\ell_v}_\sigma(\mathbf{k})-\sum_{\mathbf{q}}w_{\ell_c\ell_v}(\mathbf{q})\psi_{\sigma}(\mathbf{k}+\mathbf{q})\nonumber\\ 
    &&=E^{\ell_c\ell_v}_{\sigma}\psi(\mathbf{k}),
\end{eqnarray}
where we adopted the Keldysh-like potential in the CrSBr bilayer. We consider the mass anisotropies in the dispersion
\begin{eqnarray}
    \varepsilon_{c}(\mathbf{k})&&=\frac{k_x^2}{2m_{cx}}+\frac{k_y^2}{2m_{cy}},\\     \varepsilon_{v}(\mathbf{k})&&=-\frac{k_x^2}{2m_{vx}}-\frac{k_y^2}{2m_{vy}}\nonumber.
\end{eqnarray}
The conduction band masses are $m_{cx}=7.31m_0$, $m_{cy}=0.14m_0$ and the valence band mass are $m_{vx}=2.84m_0$, $m_{vy}=0.45m_0$\cite{klein2023bulk}.

In CrSBr bilayer, we model the screened Coulomb interaction 
as\cite{Danovich:PRB97-2018} 
\begin{equation}
v_{\ell\ell'}(\mathbf{q})=\frac{2\pi}{\epsilon q}
\frac{\kappa_{\ell\ell'}(q)}{(1+r_{\ast} q)^2-r_{\ast}^2q^2\mathrm{e}^{-2qd}} ,
\label{eqn:Keldysh}
\end{equation}
where $d=7.93$\AA \cite{LopezPaz2022} 
is the interlayer distance of the bilayer CrSBr $\kappa_{12}(q)=\kappa_{21}(q)=\mathrm{e}^{-qd}$ and $\kappa_{11}(q)=\kappa_{22}(q)=1+r_\ast q(1-\mathrm{e}^{-2qd})$. The screening length is\cite{Keldysh1979,Rytova:1967,Cudazzo2011,Ceferino2020} 
\begin{align}
    r_\ast=\frac{\epsilon_{s}-1}{\epsilon}d ,
\end{align}
where $\epsilon$ is the dielectric constant of the environment and $\epsilon_{s}$ is the dielectric constant of CrSBr. Here, we let $\epsilon_s\sim6$ which gives binding energy $E^{11}_\sigma=E^{22}_{\sigma}\approx537$~meV in vacuum ($\epsilon$=1). In this calculation, we ignore the anisotropic screening for simplicity. We note that anisotropy can be accounted for by the first principle calculations. However, we do not target purely numerical modeling and aim to describe trends, hence justifying the omission. 

In anti-ferromagnetic (AFM) phase, interlayer tunneling is not allowed. However, in the ferromagnetic (FM) phase, the intralayer and interlayer excitons hybridized due to interlayer electrons tunneling. To account for the hybridization, we can solve for the coefficients $C^{\ell\ell'}_{\sigma}$ \cite{Song:JunsarXiv-2024} 
as
\begin{align}
\begin{bmatrix}
    E^{11}_{\sigma} & -T_v & T_c &0\\
    -T_v^\ast & E^{12}_{\sigma} &0& T_c\\
    T_c^\ast & 0 & E^{21}_{\sigma}&-T_v\\
    0 & T_c^\ast & -T_v^\ast &E^{22}_{\sigma}
\end{bmatrix}\begin{bmatrix}
    C^{11}_{\sigma}\\C^{12}_{\sigma}\\C^{21}_{\sigma}\\C^{22}_{\sigma}
\end{bmatrix}=E_{b}
\begin{bmatrix}
    C^{11}_{\sigma}\\C^{12}_{\sigma}\\C^{21}_{\sigma}\\C^{22}_{\sigma}
\end{bmatrix},
\end{align}
where the transition matrix elements are
\begin{eqnarray}
T_v&&=t_v\sum_{\mathbf{k}}\bar{\psi}^{11}_\sigma(\mathbf{k})\psi^{12}_\sigma(\mathbf{k}),\\
T_c&&=t_c\sum_{\mathbf{k}}\bar{\psi}^{11}_\sigma(\mathbf{k})\psi^{12}_\sigma(\mathbf{k}). \nonumber
\end{eqnarray}

Here, we consider the relevant valence band interlayer hopping $t_v=t=12$~meV and $t_c=0$.\cite{klein2023bulk}

In our analysis, we concentrate on 1$s$ states and set the total momentum of the exciton be $\mathbf{Q}=0$ in the scattering processes, such that we characterize the low-energy exciton-exciton (X-X) interactions with elastic scattering for $\mathbf{Q}=0$ only. The X-X interaction between exciton can be calculated from the total energy of the two-exciton state, $\Omega_{\sigma}=\langle0|X_{\sigma}X_{\sigma}\mathcal{H}X^{\dagger}_{\sigma}X^{\dagger}_{\sigma}|0\rangle=2E_b+\Delta_{\sigma}$. 
The interacting potential energy is given by
\begin{align}
\Delta_{\sigma}=-2\sum_{ss'}\sum_{\tilde{s}\tilde{s}'}&\bar{C}^{\sigma}_{s'}\bar{C}^{\sigma}_{\tilde{s}'}C^{\sigma}_{s}C^{\sigma}_{\tilde{s}}
V_{s\tilde{s}}^{s'\tilde{s}'},
\end{align}
where $s=(\ell_c,\ell_v)$ is the layer double index. The exchange interaction reads
\begin{eqnarray}
V_{s\tilde{s}}^{s'\tilde{s}'}=&&\frac{1}{2A}\sum_{\mathbf{k}\tilde{\mathbf{k}}\mathbf{q}}\sum_{\ell\ell'}f^{\ell}_{s\sigma}(\mathbf{k},\mathbf{q}) w_{\ell\ell'}(\mathbf{q}) f^{\ell'}_{\tilde{s}\sigma}(\mathbf{k},-\mathbf{q})\\ &&\times \psi^{\ast}_{s'}(\mathbf{k})\psi^{\ast}_{\tilde{s}'}(\tilde{\mathbf{k}})\delta_{\tilde{\ell}_c\ell_c'}\delta_{\tilde{\ell}_v\tilde{\ell}_v'}\delta_{\ell_c\tilde{\ell}_c'}\delta_{\ell_v\ell_v'}\delta_{\mathbf{q},\mathbf{k}-\tilde{\mathbf{k}}} \nonumber
\end{eqnarray}
where $A$ is the area of the sample. The above equation gives 
\begin{equation}
    g_\mathrm{X}=A\Delta_\sigma .
\end{equation}
We note that the direct interaction vanishes since we let the total momentum of the exciton $Q=0$. Here, $\mathbf{q}$ is the transferred momentum between excitons, 
with excitonic wavefunction being expressed in $s$-index notation as $\psi_{s\sigma}(\mathbf{k})=\psi^{\ell_c\ell_v}_{\sigma}(\mathbf{k})$, and the factor
$f^{\ell}_{s\sigma}(\mathbf{k},\mathbf{q})=\delta_{\ell_c\ell}\psi_{s\sigma}(\mathbf{k}-\mathbf{q})-\delta_{\ell_v\ell}\psi_{s\sigma}(\mathbf{k})$. Using the wavefunction and $t=12$~meV, we find that the non-linearity is $g_\mathrm{X}\approx0.29$ $\mu$eV $\mu$m$^2$. Interlayer hybridization only leads to a difference within $0.01$ $\mu$eV $\mu$m$^2$.

Furthermore, the exciton-exciton interaction leads to a weak non-linear blueshift which is not sufficient to account for the non-linear shift in the experiment, particularly the redshift in AFM phase. Even though this result is from a bilayer system, we do not expect our conclusions to change significantly in a system with large number of layers (bulk). Therefore, we consider additional contribution to non-linear energy shift from the coupling with magnons.

\begin{figure}[t]
\includegraphics[width=0.9\columnwidth]{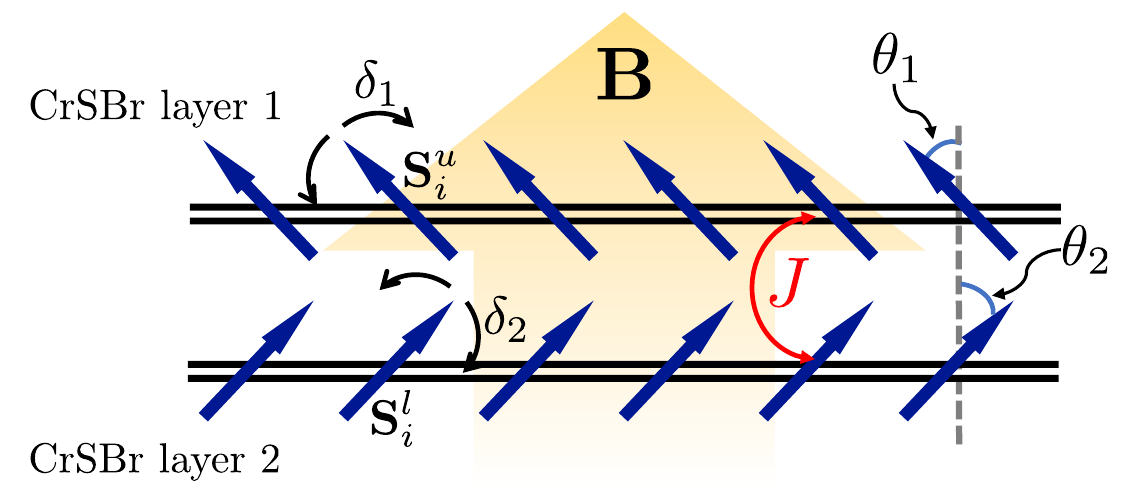}
  \caption{\textbf{Schematic diagram for the spin model in CrSBr.} The spin in the upper and lower layer at site $i$ are $\mathbf{S}^u_i$ and $\mathbf{S}^l_i$. Their interlayer spin exchange coupling is $J$. In the presence of the uniform out-of-plane magnetic field $\mathbf{B}$, the spins $\mathbf{S}^u_i$ ($\mathbf{S}^l_i$) become tilted, forming an angle $\theta_1$ ($\theta_2$) with respect to $\mathbf{B}$-field. In non-zero temperature, the spins $\mathbf{S}^u_i$ ($\mathbf{S}^l_i$) fluctuate around its equilibrium orientation with small angle $\delta_1$ ($\delta_2$).}
    \label{figure_S6}
\end{figure}

\subsection{Magnon}

In this subsection, we investigate the non-linear energy shift due to exciton-magnon coupling. From Eq.~\eqref{eqn:EX}, the magnetic spin couple to the exciton through the bandgap term $\varepsilon_{g}(\theta_1-\theta_2)$. Note that here as a magnon mode we consider spin waves at $k=0$ (Kittel mode corresponding to the uniform precession). We consider that coupling to distinct modes with non-zero momentum is suppressed due to selection rules and significant exciton broadening.

In experiment, an out-of-plane magnetic field is applied to the sample (Fig.~\ref{figure_S6}). This points the spin at the equilibrium directions defined by $\theta_{1\ast}$ and $\theta_{2\ast}$. To obtain this equilibrium angles, we model the CrSBr as a bilayer spin system with energy density (energy per unit cell) as 
\begin{align}
    E_M=&\frac{1}{N_s}\Big[\sum_{ i=1}^{N_s}J\mathbf{S}^{l}_i\cdot\mathbf{S}^{u}_i-\sum_{i=1}^{N_s}\mu_0(\mathbf{S}^{l}_i+\mathbf{S}^{u}_i)\cdot\mathbf{B}\Big]\\ \nonumber
    &-\frac{1} {N_s}\sum_{i=1}^{N_s}\Big(A_xS^{l}_{ix}S^{l}_{ix}+A_xS^{u}_{ix}S^{u}_{ix}\Big)\\ \nonumber
    &-\frac{1} {N_s}\sum_{i=1}^{N_s}\Big(A_zS^{l}_{iz}S^{l}_{iz}+A_zS^{u}_{iz}S^{u}_{iz}\Big)\Big]
\end{align}
where $\mathbf{S}^{l,u}_i$ is the spin for lower and upper layer. The total number of unit cell is $N_s$. The interlayer magnetic exchange coupling $J=24.8~\mu$eV and the anistropic exchange to the easy axis is $A_x=72.5\mu$eV and to the out-of-plane axis is $A_z=14.4~\mu$eV,\cite{Wang:PRB108-2023} where we set the hard axis anisotropic exchanged be zero. The last term is the magnetic anisotropy that gives the preferential direction of the spin in $x$-direction (we remind that CrSBr is a quasi-1D system). Here, we note that the interlayer exchange interaction $J$ has a similar strength as the anisotropy $A_x$. As a function of orientation angles the energy can be written as
\begin{align}
    E_M(\theta_1,\theta_2)=&JS^2\cos(\theta_1-\theta_2)-\mu_0SB(\cos\theta_1+\cos\theta_2)\notag\\
    &-A_xS^2(\sin^2\theta_1+\sin^2\theta_2)\\ \nonumber
    &-A_zS^2(\cos^2\theta_1+\cos^2\theta_2) ,
\end{align}
where $S=3/2$ is the spin at chromium site. The angle between the upper (lower) layer spin and the magnetic field is defined as $\theta_1$ ($\theta_2$).

To find the tilted angle with the applied magnetic field $B$, we minimize the total energy by solving
\begin{align}
    \left.\frac{\partial E_M}{\partial \theta_1}\right|_{\theta_{1\ast},\theta_{2\ast}}=0,\quad \left.\frac{\partial E_M}{\partial \theta_2}\right|_{\theta_{1\ast},\theta_{2\ast}}=0 .
\end{align}
Assuming that $\theta_1=\theta_{1\ast}$ and $\theta_2=\theta_{2\ast}$ admit minimum solution of $E_M$, this leads to
\begin{align}
    \frac{\partial E_M}{\partial \theta_1}=&-JS^2\sin(\theta_{1\ast}-\theta_{2\ast})+\mu_0SB\sin\theta_{1\ast} \nonumber\\
    &-(A_x-A_z)S^2\sin2\theta_{1\ast}=0,  \label{eqn:vartheta}\\
    \frac{\partial E_M}{\partial \theta_2}=&JS^2\sin(\theta_{1\ast}-\theta_{2\ast})+\mu_0SB\sin\theta_{2\ast} \nonumber\\
    &-(A_x-A_z)S^2\sin2\theta_{2\ast}=0 .\label{eqn:varphi}
\end{align}
Solving the above equations, we obtain the solution that minimizes $E_M$.  The saturation field $B_{\mathrm{sat}}$ can be obtained using Eqs.~\eqref{eqn:vartheta} and \eqref{eqn:varphi},
\begin{equation}
    B=\frac{2JS\sin(\theta_{1\ast}-\theta_{2\ast})+(A_x-A_z)S(\sin2\theta_{1\ast}-\sin2\theta_{2\ast})}{\mu_0(\sin\theta_{1\ast}-\sin\theta_{2\ast})}
\end{equation}
by taking the limits $\theta_1,\theta_2\to 0$. This gives the saturation magnetic field 
\begin{equation}
    B_{\mathrm{sat}}=2S\frac{(J+A_x-A_z)}{\mu_0}
\end{equation}
in the ferromagnetic phase.

The spins in CrSBr are dynamic and can fluctuate around the equilibrium directions with small angle $\delta_1$ and $\delta_2$ (see Fig. \ref{figure_S6}). We expand the energy in the vicinity of this point as
\begin{align}\label{eqn:EM}
E_M(\theta_{1\ast}+\delta_1,\theta_{2\ast}+\delta_2)\approx E_{M}(\theta_{1\ast},\theta_{2\ast})+a\delta_1^2+2b\delta_{1}\delta_{2}+c\delta_{2}^2 ,
\end{align}
where 
\begin{widetext}
\begin{align*}
    a=&\frac{\partial^2E_{M}(\theta_{1\ast},\theta_{2\ast})}{\partial\theta_{\ast1}^2}=-JS^2\cos(\theta_{1\ast}-\theta_{2\ast})+\mu_0SB\cos\theta_{1\ast}-2(A_x-A_z)S^2\cos2\theta_{1\ast},
    \\
    b=&\frac{\partial^2E_{M}(\theta_{1\ast},\theta_{2\ast})}{\partial\theta_{\ast1}\partial\theta_{\ast2}}=JS^2\cos(\theta_{1\ast}-\theta_{2\ast}),
    \\
    c=&\frac{\partial^2E_{M}(\theta_{1\ast},\theta_{2\ast})}{\partial\theta_{\ast2}^2}=-JS^2\cos(\theta_{1\ast}-\theta_{2\ast})+\mu_0SB\cos\theta_{2\ast}-2(A_x-A_z)S^2\cos2\theta_{2\ast} .
\end{align*}
\end{widetext}
We write Eq.~\eqref{eqn:EM} into the magnon normal modes as
\begin{align}
    E_M(\theta_{1\ast}+\delta_1,\theta_{2\ast}+\delta_2)\approx E_{M}(\theta_{1\ast},\theta_{2\ast})+\omega_-\eta_-^2+\omega_+\eta_+^2 ,
\end{align}
where the normal mode frequencies $\omega_{+}$ and $\omega_{-}$ are the eigenvalues of the matrix,
\begin{align}\label{eqn:omega}
    \omega_{\pm}=\frac{(a+c)\pm\sqrt{(a-c)^2+4b^2}}{2},
\end{align}
and the normal eigenmodes are

\begin{align}\label{eqn:eta}
    \eta_\pm=\frac{(\omega_{\pm}-c)}{\sqrt{b^2+(\omega_{\pm}-c)^2}}\delta_1+\frac{b}{\sqrt{b^2+(\omega_{\pm}-c)^2}}\delta_2 .
\end{align}
Therefore, the bandgap changes as 
\begin{align}
&\varepsilon_g(\theta_{1\ast}+\delta_1,\theta_{2\ast}+\delta_2)
    =\varepsilon_0-t\Big[\cos^2\frac{1}{2}\theta_{\ast}\nonumber\\&-\sin\theta_{\ast}(\beta_+\eta_+-\beta_-\eta_-)-\cos\theta_{\ast}(\beta_+\eta_+-\beta_-\eta_-)^2\Big]\label{eqn:M-X-coupling}
\end{align}
with 
\begin{align}\label{eqn:beta}
    \beta_+=\frac{(1+(\omega_--c)/b)\alpha_+}{\omega_+-\omega_-},\quad    \beta_-=\frac{(1+(\omega_+-c)/b)\alpha_-}{\omega_+-\omega_-}
\end{align}
The magnon-exciton coupling is zero at AFM ($\theta_{1\ast}=-\theta_{2\ast}=\pi/2$) and FM ($\theta_{1\ast}=\theta_{2\ast}=0$) phase. This implies there is no redshift in this phase if we disregard the fluctuation of the quadratic terms.

\subsection{Thermal effect and incoherent magnon}

In nonzero finite temperature, we measure the average exciton energy in Eq.~\eqref{eqn:EX} due to the thermal fluctuation of the spins.\cite{dirnberger2023magneto}
\begin{align}
\bar{E}_X(\theta_{1\ast},\theta_{2\ast})=&\bar{\varepsilon}_g(\theta_{1\ast}-\theta_{2\ast})+E_b+\rho_\mathrm{X} g_\mathrm{X},
\label{eqn:EX-th}
\end{align}
where the average is
\begin{align}
    \bar{\varepsilon}_g(\theta_{1\ast}-\theta_{2\ast})
    &=\varepsilon_0-t\Big[\cos^2\frac{1}{2}\theta_{\ast}-\sin\theta_{\ast}(\beta_+\langle\eta_+\rangle-\beta_-\langle\eta_-\rangle)\nonumber \\&-\cos\theta_{\ast}(\beta_+^2\langle\eta_+^2\rangle-2\beta_+\beta_-\langle\eta_+\eta_-\rangle+\beta_-^2\langle\eta_+^2\rangle)\Big] .
\end{align}
The small fluctuation around $\theta_{\ast}$, denoted as $\eta_{\pm}$, can take positive and negative values. Therefore, we may expect $\langle\eta_\pm\rangle=0$ and disregard the linear coupling term (second term). However, in the last term, we expect $\langle\eta_\pm^2\rangle\propto n_{\pm}$ where $n_{\pm}$ is the total number of the ($\pm$) magnons modes in the sample. $\eta_\pm^2$ is proportional to the amplitude square. For the cross-term, we have $\langle\eta_+\eta_-\rangle=\langle\eta_+\rangle\langle\eta_-\rangle=0$, since $\eta_+$ and $\eta_-$ are two independent orthogonal modes. 
This reduces the bandgap to
\begin{align}
    \bar{\varepsilon}_g(\theta_{1\ast},\theta_{2\ast})
    =&\varepsilon_0-t\Big[\cos^2\frac{1}{2}\theta_{\ast}-\cos\theta_{\ast}(\beta_+^2\langle\eta_+^2\rangle+\beta_-^2\langle\eta_+^2\rangle)\Big]
\end{align}
To calculate the $\langle\eta_\pm\rangle$, we model the thermal effects using canonical ensemble with the partition function as
\begin{align}
    Z&=\int_{-\eta_+^c}^{\eta_+^c} d\eta_+\int_{-\eta_-^c}^{\eta_-^c} d\eta_-\mathrm{e}^{-\frac{1}{k_BT} [E_{M}(\theta_{1\ast},\theta_{2\ast})+\omega_-\eta_-^2+\omega_+\eta_+^2]}\\
    &=\mathrm{e}^{-\frac{1}{k_BT} E_{M}(\theta_{1\ast},\theta_{2\ast})}\prod_{i=\pm}\sqrt{\frac{\pi k_BT}{\omega_i}}\text{erf}\Big(\eta^c_i\sqrt{\frac{\omega_i}{k_BT}}\Big) \nonumber,
\end{align}
where $\eta_\pm^c$ is the cutoff that are related to the maximum fluctuation in $\delta_1$ and $\delta_2$. Therefore, this gives
\begin{align}\label{eqn:all_w}
    \langle\eta_\pm^2\rangle=&\frac{1}{Z}\int_{-\eta_+^c}^{\eta_+^c} d\eta_+\int_{-\eta_-^c}^{\eta_-^c} d\eta_-\eta_\pm^2\mathrm{e}^{-\frac{1}{k_BT} [E_{M}(\theta_{1\ast},\theta_{2\ast})+\omega_-\eta_-^2+\omega_+\eta_+^2]}\notag\\
    =&\Big(\frac{k_BT}{2\omega_{\pm}}\text{erf}(\eta_\pm^c\sqrt{\frac{\omega_\pm}{k_BT}})-\frac{\eta_\pm^c}{\sqrt{\pi}}\mathrm{e}^{-\frac{(\eta_\pm^c)^2}{k_BT}}\Big)/\text{erf}(\eta_\pm^c\sqrt{\frac{\omega_\pm}{k_BT}})
\end{align}
In the low-temperature $\omega_{\pm}/k_BT\to\infty$. This gives the following simple result,
\begin{align}\label{eqn:finite_w}
    \langle\eta_\pm^2\rangle=\frac{1}{2}k_BT/\omega_{\pm} .
\end{align}
Here, the temperature $T$ is the magnonic temperature which is proportional to the pump intensity. This result can also be understood intuitively by considering the total number of thermally-excited magnons, since the $k_BT$ is the thermal energy and $\omega_\pm$ is a single magnon energy. 

Note that the result in Eq. \eqref{eqn:finite_w} holds only for the case of magnon energy being sufficiently large. However, for the cases where the magnon energy is small we can no longer take $\omega_{\pm}/k_BT\to\infty$. In the case we take a limit $\omega_\pm\to 0$ in Eq.\eqref{eqn:all_w}, leading to
\begin{align}\label{eqn:zero_w}
    \langle\eta_\pm^2\rangle\approx\frac{(\eta_\pm^c)^2}{3} .
\end{align}
In this case, almost all available low-energy magnons get excited.

Therefore, we arrive at the exciton energy written as

\begin{align}
    &\bar{E}_X(\theta_{1\ast},\theta_{2\ast})
    =\varepsilon_0+E_b+\rho_\mathrm{X} g_\mathrm{X}\nonumber\\&-t\Big[\cos^2\frac{1}{2}(\theta_{1\ast}-\theta_{2\ast})-\Big(\beta_+^2\langle\eta_+^2\rangle+\beta_-^2\langle\eta_-^2\rangle\Big)\cos(\theta_{1\ast}-\theta_{2\ast})\Big]\label{eqn:EX-T}
\end{align}

where $\varepsilon_g(\theta_{1\ast}-\theta_{2\ast})$ is the bandgap, $E_b$ is the exciton binding energy, and $g_\mathrm{X}$ is the exciton-exciton exchange interacting strength. Using Eqs. \eqref{eqn:omega}, \eqref{eqn:eta}, and \eqref{eqn:beta}, we obtain the important result for explaining the findings in different magnetic configurations:
\begin{align}
    &\bar{E}_X(\pi/2,-\pi/2)
    =\varepsilon_0-t\frac{k_BT}{\omega_-}+E_b+\rho_\mathrm{X} g_\mathrm{X}, \text{(AFM)}\\
    &\bar{E}_X(0,0)
    =\varepsilon_0-t+t\frac{2}{3}(\eta_-^c)^2+E_b+\rho_\mathrm{X} g_\mathrm{X}, \text{(FM)}
\end{align}
where in the FM phase the magnon energy near $B=B_{\mathrm{sat}}$ is very small. As $B>B_{\mathrm{sat}}$, the incoherent magnon ($\eta_-$-mode) shift in FM phase take the general form in Eq.~\eqref{eqn:all_w}.

To investigate the polaritonic response in CrSBr, we recognize that the created exciton is hybridized with the intrinsic cavity mode forming a self-hybridize polariton in strong light-matter coupling regime. We model this self-hybridized polariton as follows:
\begin{equation}
    \mathcal{H}_p=\begin{bmatrix}
        \omega_c&\Omega\\
        \Omega & \bar{E}_X(\theta_{1\ast},\theta_{2\ast})
    \end{bmatrix} ,
\end{equation}
where $\omega_c$ is the intrinsic cavity mode with $\omega_c\approx \varepsilon_{0}+E_b$. The Rabi splitting $\Omega$ also experiences a non-linear response due to phase space filling effect.\cite{Song:PRR6-2024} The lower polariton in this system has the energy
\begin{equation}
    \mathcal{E}=\frac{1}{2}[\omega_c+E_X(\theta_{1\ast},\theta_{2\ast})]-\sqrt{\frac{1}{2}[\omega_c-E_X(\theta_{1\ast},\theta_{2\ast})]^2+\Omega^2}
\end{equation}
In the expression above, we approximate the Rabi splitting as $\Omega=\Omega_0\left(1-\frac{1}{2}\frac{a_X^2\rho_\mathrm{X}}{(1+\gamma_c)(1+\gamma_v)}\right)$,\cite{Song:PRR6-2024} where $\gamma_{c/v}=m_{c/v}/(m_c+m_v)(a_X/\sqrt{A})$ with $m_{c/v}=\sqrt{m_{c/v,x}m_{c/v,y}}$ being the geometrical average of the conduction/valence band masses and $a_X=\langle r\rangle$ being the average distance between the electron and the hole (exciton size) that from theoretical estimation is
\begin{align}
    a_X=\begin{cases}
        1.0\text{nm}& \text{AFM, intralayer exciton,}\\
        1.1\text{nm}& \text{FM, hybridized exciton.}
    \end{cases}
\end{align}
Also, the Rabi splitting in low-density regime is $\Omega_0\approx0.24$~eV.\cite{dirnberger2023magneto} 
We write the energy shift due to the small change in temperature $\Delta T$ and exciton density $\Delta \rho_\mathrm{X}$ by expanding it as 
\begin{align}
    \Delta\mathcal{E}=&\mathcal{B}\Delta T+\mathcal{A}\Delta \rho_\mathrm{X},
\end{align}
where
\begin{align}
    \mathcal{B}=&\Big[\frac{1}{2}+\frac{\omega_c-\bar{E}_X^{(0)}}{2\Lambda}\Big]\frac{\partial \bar{E}_X}{\partial T}, \\
     \mathcal{A}=&\Big(\frac{1}{2}+\frac{\omega_c-\bar{E}_X^{(0)}}{2\Lambda}\Big)\frac{\partial \bar{E}_X}{\partial \rho_\mathrm{X}}-\frac{\Omega_0}{2\Lambda}\frac{\partial \Omega}{\partial \rho_\mathrm{X}}.
\end{align}
Here, we have defined $\Lambda=\sqrt{\frac{1}{2}(\omega_c-E_X^{(0)})^2+\Omega_0^2}$ with $E_X^{(0)}=\varepsilon_0-t\cos^2\frac{1}{2}(\theta_{1\ast}-\theta_{2\ast})+E_b$. The derivatives are
\begin{align}
    \frac{\partial \bar{E}_X}{\partial T}=&\Big(\beta_+^2\frac{\partial\langle \eta_+^2\rangle}{\partial T}+\beta_-^2\frac{\partial\langle \eta_-^2\rangle}{\partial T}\Big)\cos\tfrac{1}{2}(\theta_{1\ast}-\theta_{2\ast}),\\
    \frac{\partial \bar{E}_X}{\partial \rho_\mathrm{X}}=&g_\mathrm{X},\\
    \frac{\partial \Omega}{\partial \rho_\mathrm{X}}=&-\frac{\Omega_0a_X^2/2}{(1+\gamma_c^2)(1+\gamma_v^2)}.
\end{align}
Assuming $\gamma_{c,v}\approx0$, we find the saturation factor $\frac{\partial \Omega}{\partial \rho_\mathrm{X}}\approx-0.24$ $\mu$eV$\mu$m$^2$ which is as large as $g_\mathrm{X}$ and it is another important nonlinear effect.
The energy shift due to laser power is
\begin{widetext}
\begin{align}
    \Delta\mathcal{E}_{\mathrm{AFM}}=&-\Big[\frac{1}{2}+\frac{\omega_c-\bar{E}_X^{(0)}}{2\Lambda}\Big]\frac{tk_B}{\omega_-}\Delta T +\Big[\Big(\frac{1}{2}+\frac{\omega_c-\bar{E}_X^{(0)}}{2\Lambda}\Big)g_\mathrm{X}+\frac{\Omega_0}{2\Lambda}\frac{a_X^2/2}{(1+\gamma_c^2)(1+\gamma_v^2)}\Big]\Delta \rho_\mathrm{X}, &\text{(AFM)}\\
    \Delta\mathcal{E}_{\mathrm{FM}}=& \Big[\Big(\frac{1}{2}+\frac{\omega_c-\bar{E}_X^{(0)}}{2\Lambda}\Big)g_\mathrm{X}+\frac{\Omega_0}{2\Lambda}\frac{a_X^2/2}{(1+\gamma_c^2)(1+\gamma_v^2)}\Big]\Delta \rho_\mathrm{X},&\text{(FM)}
\end{align}
\end{widetext}
where the magnon energy $\omega_-=S^2(J+A_x-A_z)$ with the magnetic exchange couplings $J=24.8$~$\mu$eV, $A_x=72.5$~$\mu$eV and $A_z=14.4$~$\mu$eV \cite{Wang:PRB108-2023}.

In the FM phase, since the magnon energy is very small, almost all the available magnonic excited states are depleted immediately with small temperature change. This results in the very low temperature-dependent blueshift. Therefore, in this case, the exciton energy nonlinear blueshift is mostly coming from the exciton-exciton exchange interaction. This is consistent to the non-linear response that we observed in high-power measurement. We found the maximum blue shift is $\Delta\mathcal{E}=+2.3$~meV in Fig.~3k of the main text.

As we can see, the magnon fluctuating term in AFM phase is negative leading to the redshift. Moreover, we find this effect is significant. Changing the temperature by $\Delta T\approx 1.6$ K is sufficient to generate a $\Delta\mathcal{E}_{\mathrm{FM}}=-2$~meV redshift in exciton energy if we assume the maximum blueshift coming from the saturation and exciton-exciton interaction is $\Delta\mathcal{E}_{\mathrm{FM}}=+2.3$~meV. Furthermore, we have let $\omega_c-E^{(0)}_X\approx0$ (small detuning). Therefore, the blueshift coming from exciton-exciton interaction is compensated, leading to the overall redshift.

\section{Theoretical treatment of second order quantum correlations}
\label{ap:g2mc}

In order to simulate the second order correlation theoretically, we apply the Lindblad master equation to describe the dissipative nature of the polariton system.
Under the Born approximation, we decomposed the total density matrix operator as $\rho_{tot} =\rho \otimes \rho_{pump} \otimes \rho_{ph}$, which $\rho$, $\rho_{pump}$ and $\rho_{ph}$ are the density operators of polaritons, reservoir particles and phonons, respectively.
Partially tracing out the pumping and phonon degrees of freedom, the evolution of the density matrix is 

\begin{equation}
    \frac{d\rho}{d t} = -\frac{i}{\hbar} \left[ \hat{\mathcal{H}} ,\rho \right] + \sum_s \left( \hat{J}_s \rho \hat{J}_s^\dagger -\frac{1}{2}\lbrace \hat{J}_s^\dagger \hat{J}_s, \rho \rbrace \right)
\end{equation}

The first term is the Hamiltonian dynamics of the system in different magnetic orders
\begin{align}
    \hat{\mathcal{H}} =& \sum_i E_i \hat{a}_i^\dagger \hat{a}_i + \sum_i U_0 \hat{a}_i^\dagger \hat{a}_i^\dagger \hat{a}_i \hat{a}_i \nonumber \\
    +& \sum_i U_1 \hat{a}_i^\dagger \hat{a}_i^\dagger \hat{a}_i^\dagger \hat{a}_i^\dagger \hat{a}_i \hat{a}_i \hat{a}_i \hat{a}_i
\end{align}
where $E_i$ is the lower polariton eigenmode from Fig.~3c of the main text at linear regime, and $U_0$ and $U_1$ are the polariton-polariton scattering strength to simulate the nontrivial redshift and blueshift with different pumping intensity for the AFM and FM phases, respectively. This is the main difference between different magnetic order in our quantum treatment.

\begin{figure}[t]
    \includegraphics[width=0.9\columnwidth]{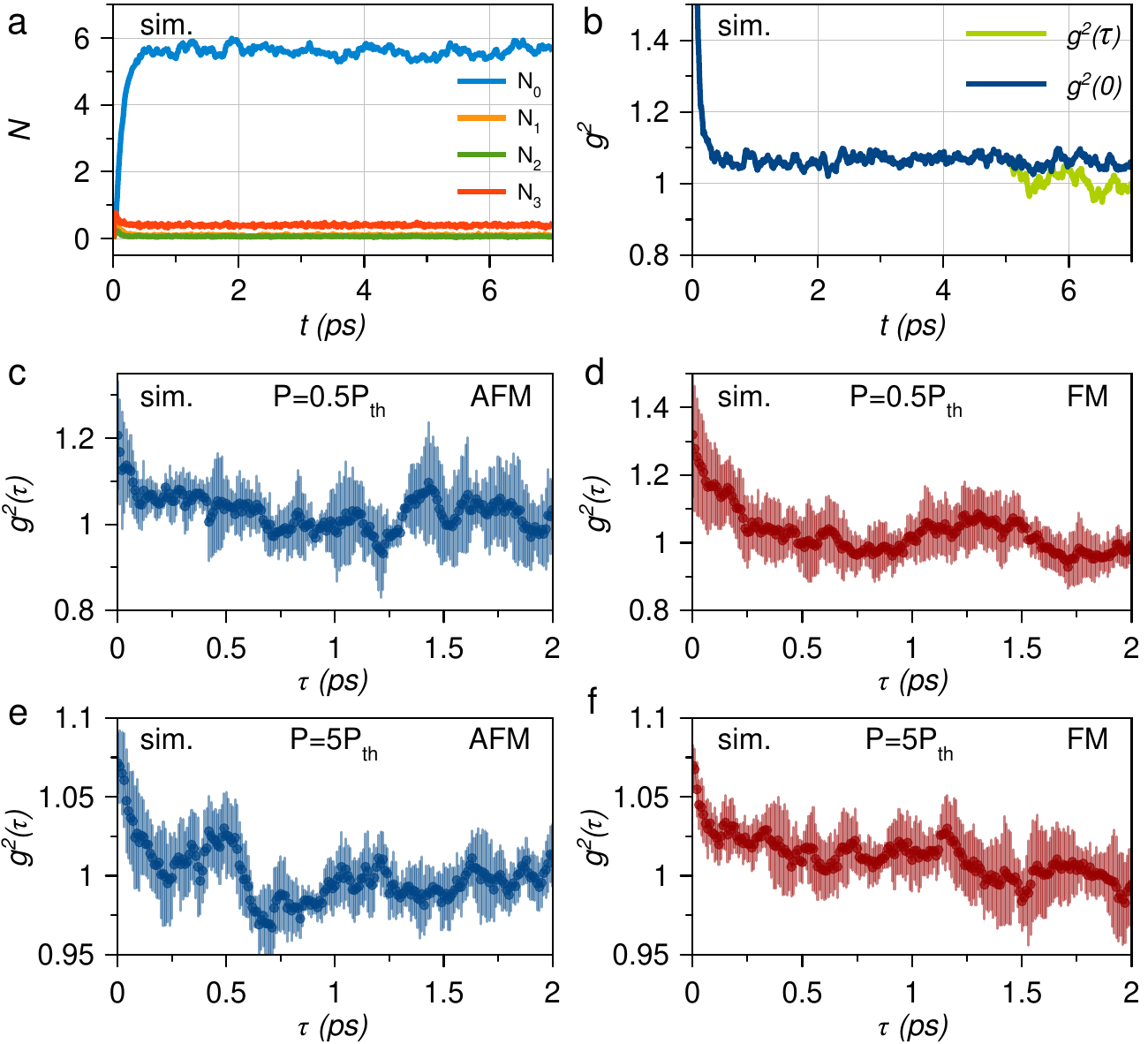}
    \caption{\textbf{Simulation of second-order correlation. }\textbf{a.} Single `measurement' of the number of particle as a function of time for different eigenmode. \textbf{b.} The result of $g^2(0)$ and $g^2(\tau)$. The number of particle and correlation function $g^2(0)$ clearly show the system reach the steady state at $t=5$ ps. \textbf{c}-\textbf{f.} are the typical averaged results of $g^2(\tau)$ and their standard derivation averaged over 5 independent measurements at two different pumping intensity and two magnetic orders as a function of delay time $\tau$.}
    \label{fig:figure_S10}
\end{figure}

The second term stands for the incoherent process, it represents the summation of all the dissipative channels \textit{"i"} associated with the dissipative operator $\hat{J}_i$.
Applying the quantum treatment \cite{flayac2015} for polariton system, we can get the jumping operators for pumping channel,
\begin{align}
    \hat{J}^+_i =& \sqrt{\gamma_i \bar{n}_p \left(E\right)} \hat{a}_i^\dagger \\
    \hat{J}^-_i =& \sqrt{\gamma_i \left[ \bar{n}_p \left(E\right)+1 \right]} \hat{a}_i
\end{align}
where $\gamma_i$ is the decay rate of polaritons, $\bar{n}_p \left(E\right)$ is the Bose distribution to describe the statistics of reservoir particles and the pumping intensity onto a specific mode is then described by $P_i=\gamma_i \bar{n}_p \left(E_i \right)$. 
Similarly, the jumping operators for the polariton-phonon interaction channels are
\begin{align}
    \hat{J}^+_{ij} =& \sqrt{\gamma^{ph}_{ij} \bar{n}_{ph}\left(E_i -E_j\right)} \hat{a}_i^\dagger \hat{a}_j \\
    \hat{J}^-_{ij} =& \sqrt{\gamma^{ph}_{ij} \left[ \bar{n}_{ph}\left(E_i - E_j\right) +1 \right]} \hat{a}_i \hat{a}_j^\dagger
\end{align}
where $\gamma^{ph}_{ij}$ is the effective polariton-phonon scattering strength between different eigenmode $i$ and $j$, $\bar{n}_{ph} \left(E\right)$ is the Boson statistics for phonon reservoir.

\begin{figure}[t]
\includegraphics[width=0.5\textwidth]{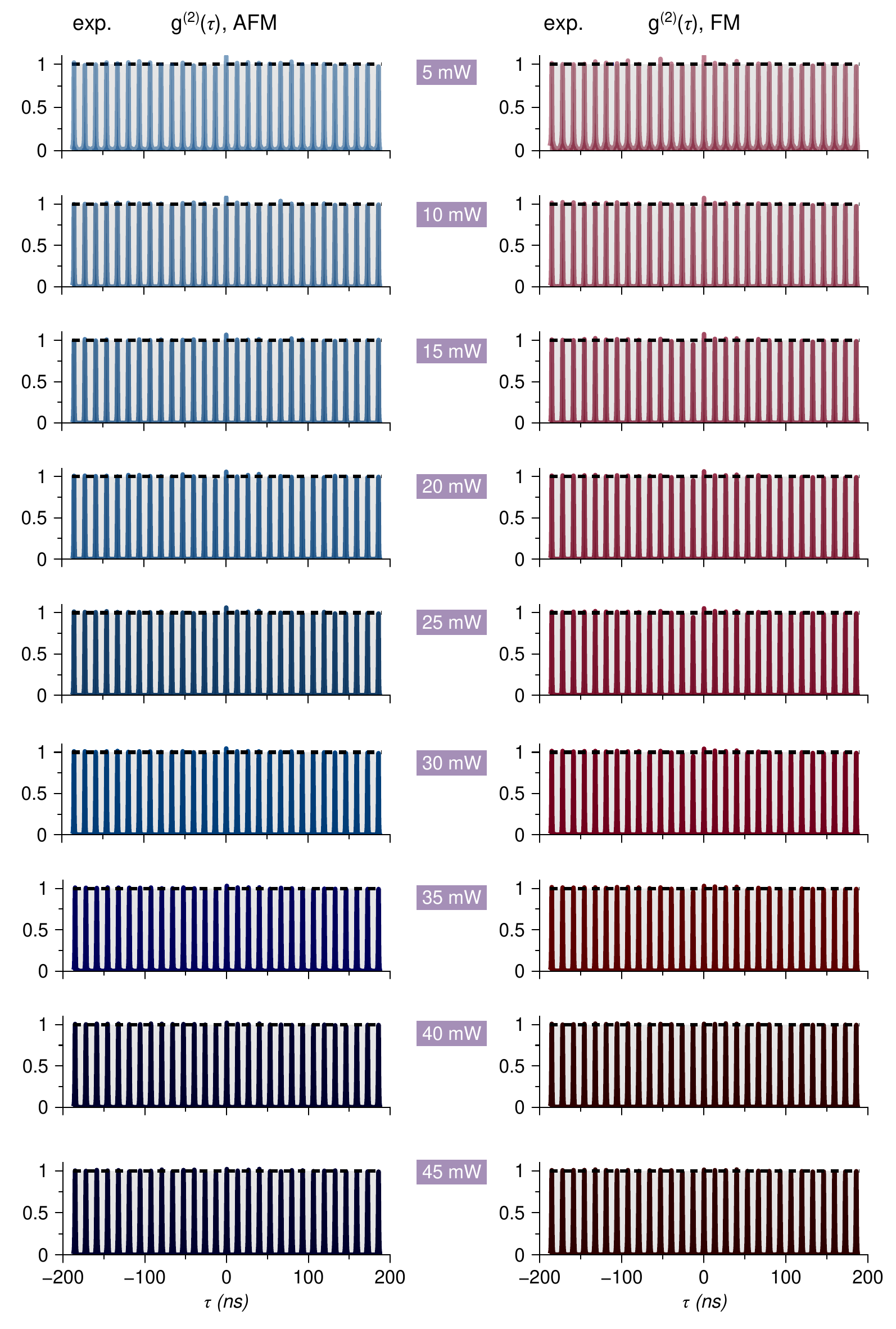}
\caption{\textbf{Pump power dependent measurements of the second-order correlation functions} ${g^2(\tau)}$ of the lower polariton branches (LPBs) with the AFM and FM orders (LPBs in Fig. 3c of the main text). The flat dashed lines at the unity mark the Poissonian level. The experimental details can be found in Methods part of the main text. The auto-correlation $\overline{g^2(0)}$ monotonously decreases from $\sim$1.1 towards 1. $\overline{g^2(0)}$ at different pump powers and magnetic orders are summarized in Fig. 5c of the main text.}
\label{figure_S8}
\end{figure}

In second quantized approach, the second order temporal correlation in Eq. (2) of the main text converts to
\begin{equation}
    g^{(2)}\left(\tau\right) = \frac{\langle \hat{a}^\dagger(t) \hat{a}^\dagger(t+\tau) \hat{a}(t+\tau) \hat{a}(t)\rangle}{\langle \hat{a}^\dagger(t) \hat{a}(t) \rangle \langle \hat{a}^\dagger(t+\tau) \hat{a}(t+\tau) \rangle}.
\end{equation}
By employing the model described above and using quantum Monte Carlo approach, we can numerically investigate the second-order temporal correlation $g^{(2)}$ of the ground state.
In each `measurement', we first let the system evolve to time $t$ which is the steady state for the system and then start the calculation of $g^2(\tau)$.
In our simulation, the pumping threshold $P_{th}$ is defined such that beyond this pumping intensity, the polaritons start to be macroscopically occupied.

The numerical calculation details are the following.
In each simulation, we consider $4$ polariton eigenstates.
The maximum number of excitation is $N_{max}=12$ for the ground state and $N_{max}=3$ for other eigenstate.
Assuming the pumping can only affect the highest eigenstate, we set $P_i=0$ for other eigenmode.
Our numerical result is the averaged result of $6$ independent `measurement'.
In each `measurement', to reach a steady state from the empty initial state, we first evolve the system up to $t=7$     ps by quantum Monte Carlo approach with $N_1=500$ trajectories.

Then, in this specific `measurement', for each trajectories,  we calculate the $g^2(\tau)$ by another independent quantum Monte Carlo approach with $N_2=500$ trajectories at $t=5$ ps.
Thus, in each `measurement', the result of $g^2(\tau)$ is averaged over $N_1\times N_2=250000$ trajectories in total. 
In Figs.~\ref{fig:figure_S10}a and \ref{fig:figure_S10}b, we show the typical result for a single `measurement'. Figs. \ref{fig:figure_S10}c-\ref{fig:figure_S10}f shows the typical averaged $g^2(\tau)$ for two magnetic orders with two pumping intensities.
At last, we summarize the parameters used in our simulation in Table.~\ref{tab:t-2}.

\begin{widetext}
\begin{center}
\begin{table}[ht]
\begin{tabular}{|c|c|c|c|c|c|c|}
    \hline
     Phase & $\lbrace E_i\rbrace$ (meV) & $\gamma_i$ [meV] & $\gamma^{ph}_i$ [meV] & $U_0$ [meV]  & $U_1$ [meV] & Phonon Temperature [K]\\
    \hline\hline 
     AFM & $\lbrace 1284.33, 1290.63, 1294.49, 1321.79\rbrace$ & $6.3$ & $11.2$ & $-0.108$ & $0.0015$ & $3$ \\
     \hline
     FM & $\lbrace 1273.23, 1278.53, 1282.09, 1309.23\rbrace$ & $6.2$ & $11.2$ & $0.0402$ & $0.000658$ & $3$\\
     \hline 
\end{tabular}
\caption{Parameters used in the second-order correlation simulation.}
\label{tab:t-2}
\end{table}
\end{center}
\end{widetext}

\bibliography{apssamp}
\end{document}